\documentclass[fleqn,usenatbib,sort]{mnras}

\usepackage{newtxtext,newtxmath}

\usepackage[T1]{fontenc}

\DeclareRobustCommand{\VAN}[3]{#2}
\let\VANthebibliography\thebibliography
\def\thebibliography{\DeclareRobustCommand{\VAN}[3]{##3}\VANthebibliography}

\usepackage{graphicx}	
\usepackage{amsmath}	

\usepackage[dvipsnames]{xcolor}
\newcommand{\rev}{}

\newcommand{\gaia}{\textit{Gaia} }
\newcommand{\kms}{$\mathrm{km\,s^{-1}}$}

\usepackage{multirow}
\usepackage{siunitx} 
\usepackage{gensymb}
\usepackage{subfigure}
\usepackage{longtable}
\usepackage{lipsum,pdflscape}

\title[HVSs in the Galactic bulge]{New candidate hypervelocity red clump stars in the inner Galactic bulge}

\author[A. Luna et al.]{
A.\,Luna,$^{1,2}$\thanks{E-mail: alonso.luna@eso.org}
T.\,Marchetti,$^{1}$
M.\,Rejkuba,$^{1}$
N.\,W.\,C.\,Leigh,$^{3,4}$
J.\,Alonso-García,$^{5,6}$
\newauthor
A.\,Valenzuela Navarro,$^{6,7}$
D.\,Minniti$^{2,8,9}$
and L.\,C.\,Smith$^{10}$
\\
$^{1}$European Southern Observatory, Karl-Schwarzschild-Straße 2, 85748 Garching, Germany\\
$^{2}$Instituto de Astrofísica, Facultad de Ciencias Exactas, Universidad Andrés Bello, Fernández Concha 700, Las Condes, Santiago, Chile\\
$^{3}$Departamento de Astronomia, Universidad de Concepcion, Avenida Esteban Iturra Casilla 160-C, Region Bio Bio 403000, Concepción, Chile\\
$^{4}$Department of Astrophysics, American Museum of Natural History, New York, NY 10024, USA\\
$^{5}$Centro de Astronom\'ia (CITEVA), Universidad de Antofagasta, Av. Angamos 601, Antofagasta, Chile\\
$^{6}$Millennium Institute of Astrophysics, Nuncio Monseñor Sotero Sanz 100, Of. 104, 7820436 Providencia, Santiago, Chile\\
$^{7}$Instituto de Astrofísica, Facultad de Física, Pontificia Universidad Católica de Chile, Av. Vicuña Mackenna 4860, 7820436 Macul, Santiago, Chile\\
$^{8}$Vatican Observatory, V00120 Vatican City State, Italy \\
$^{9}$Departamento de Fisica, Universidade Federal de Santa Catarina, Trinidade 88040-900, Florianopolis, Brazil \\
$^{10}$Institute of Astronomy, University of Cambridge, Madingley Rise, Cambridge, CB3 0HA, UK\\
}

\date{Accepted XXX. Received YYY; in original form ZZZ}

\pubyear{2023}

\begin{document}
\label{firstpage}
\pagerange{\pageref{firstpage}--\pageref{lastpage}}
\maketitle

\begin{abstract}
We search for high-velocity stars in the inner region of the Galactic bulge using a selected sample of red clump stars. Some of those stars might be considered hypervelocity stars (HVSs). Even though the HVSs ejection relies on an interaction with the supermassive black hole (SMBH) at the centre of the Galaxy, there are no confirmed detections of HVSs in the inner region of our Galaxy. With the detection of HVSs, ejection mechanism models can be constrained by exploring the stellar dynamics in the Galactic centre through a recent stellar interaction with the SMBH. Based on a previously developed methodology by our group, we searched with a sample of preliminary data from version 2 of the Vista Variables in the Via Lactea (VVV) Infrared Astrometric Catalogue (VIRAC2) and \gaia DR3 data, including accurate optical and NIR proper motions. This search resulted in a sample of \rev{46} stars \rev{with transverse velocities larger than the local escape velocity} within the Galactic bulge, of which \rev{4} are prime candidate HVSs with high-proper motions consistent with being ejections from the Galactic centre. Adding to that, we studied a sample of reddened stars without a \gaia DR3 counterpart and found \rev{481} stars \rev{with transverse velocities larger than the local escape velocity}, from which \rev{65} stars have proper motions pointing out of the Galactic centre and are candidate HVSs. In total, we found \rev{69} candidate HVSs pointing away from the Galactic centre \rev{with transverse velocities larger than the local escape velocity}.
\end{abstract}




\section{Introduction}

The Galactic bulge is the region 2.5 kpc around the Galactic centre (GC), and within the $\Lambda$ cold dark matter paradigm, it was the first Galactic structure that formed through the hierarchical merging of less massive structures \citep{2018ARA&A..56..223B,1995MNRAS.276..549S,2005ApJ...622L...9S}. 
The environment in such region is extreme, with high stellar crowding and in its centre resides a supermassive black hole (SMBH), Sgr A*, with a mass of $4.3\times10^6\,\mathrm{M_{\odot}}$ \citep{2022A&A...657L..12G,Genzel2010,Ghez2008,LFRodriguez1978PhDT}. The high stellar density favours interactions amongst stars and between stars and SgrA*.

An interaction between a triple system can result in one of the stars acquiring a velocity larger than the local standard of rest. The identification of a star after the interaction gives insight into the dynamics surrounding the interaction itself. For example, \citet{Blaauw1961} showed that runaway stars with peculiar velocities larger than $30$ \kms can form after a binary stellar system disruption in the supernova explosion of one of its components.  \citet{Poveda1967} investigated dynamical stellar interactions during the collapse of small clusters of massive stars, noting that runaway stars can result from such interactions acquiring velocities that exceed $35$ \kms and in some cases reaching nearly 200 \kms.


Hypervelocity stars (HVSs) are the fastest stars in the Galaxy.
They were initially defined as stars ejected after a three-body interaction of a binary system with SgrA*; one star is ejected as an HVS, while the other remains attached to the SMBH as an S-Star\footnote{The stars within 0.01 pc of the GC, namely the S-cluster, 
are young B-type stars, with ages consistent with the HVSs detected and confirmed in the Galactic Halo, supporting the origin scenario of both HVSs and the S-stars.}\citep{Hills1988}.
This is known as the Hills mechanism, the most accepted and successful HVS ejection scenario, but not the only one. 
Alternative mechanisms that explain how an HVS acquires its large velocity include the interaction between a binary massive black hole (bMBH) and a single star \citep{YuTremaine2003}; or between a globular cluster and a SMBH \citep[][]{Brown2015, Capuzzo2015,fragione2017,Irrgang2019,Neunteufel2020}. In a tidal disruption event, the ejection velocity of a binary component is proportional to the binary components separation and the mass of the BH: $v_{ej}\propto (a)^{-1/2}(M/M_{\odot})^{1/6}$. Given the mass of Sgr A*, $4\times10^6\,\mathrm{M_{\odot}}$, HVSs can be ejected at up to $\sim4\,000$ \kms \citep{Rossi2021,Rossi2017,Sari2010,Hills1988}, which is larger than the escape velocity. The local escape velocity is $\sim830$ \kms in the GC, $\sim$320 \kms in the halo (50 kpc) and $\sim$530 \kms at the Sun location \citep{Rossi2017,Deason2019}.
Besides explaining the extreme velocities of HVSs, the Hills mechanism can explain the presence of young stars close to the GC, where an \textit{in situ} formation is unlikely because a molecular cloud would not survive such an extreme environment. In this scenario, the young stars close to the GC could be the remnants of a binary system disruption by Sgr A* \citep{Hills1988,2021MNRAS.501.3088G}.


Since the first discovery by \citet{Brown2005}, there are around 20 confirmed HVSs and over 500 candidates \citep{Boubert2018}.
The high stellar crowding as well as large and patchy extinction in the central regions of the Milky Way hampered the detection of HVSs close to their origin. They were identified in a more favourable environment, the Milky Way halo, with velocities ranging from 300 up to 1700 \kms, exceeding the local escape speed \citep[e.g.][]{Brown2018,Kollmeier2009,Kenyon2014,Palladino2014}. Almost all of them are B-type stars with masses between 2.5 and 4$M_{\odot}$, except for the fastest star amongst them: HVS-S5 \citep{Koposov2020}, an A-type star with a velocity of 1700 \kms. This star is also the only HVS whose travel direction and orbit points to the origin from the GC, favouring the Hills mechanism as responsible for its ejection.
For other cases, the Hills mechanism can be ruled out, as the orbits of the stars suggest an origin in the disc, in the Magellanic Clouds \citep[e.g.][]{Przybilla2008,Lennon2017, Boubert2020,Evans2021}, or the Sagittarius dwarf spheroidal galaxy \citep{Huang2021, 2022ApJ...933L..13L}. 

In a recent study, \citet{2023AJ....166...12L} combined multiple spectroscopic large surveys and found that the ejection from dwarf galaxies and globular clusters is more prominent for the metal-poor late-type halo HVS production. 
Based on \gaia data, and follow-up spectroscopic observations, \citet{2023arXiv230603914E} identified four hypervelocity white dwarfs, travelling at space velocities larger than 1\,300 \kms after the double detonation -- helium in the surface, followed by carbon in the core -- of the more massive WD in a binary system of WDs. The objects have crossed the Galactic plane on multiple occasions, getting accelerated or slowed down, but none of them have trajectories that trace back to the GC. The analysis of positions and velocity distributions of the fastest stars in proximity to the GC offers insights into the shape of the Galactic potential \citep{Kenyon2008}.





HVSs within the Galactic bulge provide additional constraints on the characteristics of the stellar population in that particular region of the Galaxy. This includes the joint constraints of the stellar initial mass function of the GC and the HVS ejection rate, and the binary separation distribution \citep[see][]{Evans2022,Marchetti2022,Rossi2014,Rossi2017}.


\citet{Luna2019} presented the first HVS candidates in the Galactic bulge using the VVV near IR survey data.
This work is extended here using the \gaia DR3 data in combination with a larger set of improved VVV-based NIR proper motions: VIRAC2. We search for HVS candidates that can be further studied spectroscopically to confirm their nature and constrain their origin.

The paper is organised as follows: in Section 2, we describe the data sets. In Section 3 we present the selection of Red Clump stars and the computation of distances and tangential velocities. The resulting sample of high-velocity stars and HVSs candidates from the VIRAC2-\textit{Gaia} DR3 crossmatch is presented in Section 4 and in Section 5 the equivalent sample of those stars having only VIRAC2 data. In Section 6, we summarise our conclusions.

\section{The \gaia DR3 and VIRAC2 catalogues}


\subsection{\gaia DR3}

The \gaia third data release (DR3)  provides, amongst other information, optical photometry ($G$, $G_{BP}$ and $G_{RP}$) and astrometry (parallaxes and proper motions) for $\sim$1.5 billion sources in the Galaxy, with 34 months of observations \citep{GaiaDR3}. \gaia DR3 doubled the precision in proper motions and improved by 30\% the precision in parallaxes with respect to \gaia DR2. In the same vein, the errors improved by a factor of $\sim$2.5 in proper motions and up to 40\% in parallaxes \citep{GaiaED32021,GaiaEDR3_Fabricius2021,Gaia_Lindegren2016,Gaia_vanLeeuwen2017}.

The survey has completeness below 60\% for sources at $G\sim19$ or fainter and in stellar densities of about $5\times10^5\,\mathrm{stars}\,\deg^{-2}$, which 
is characteristic of crowded fields such as globular clusters. Its completeness further drops below 20\% even for brighter sources in fields with stellar densities of about $10^6-10^7\,\mathrm{stars}\,\deg^{-2}$ that are found in the Galactic bulge \citep{Fabricius2021,Cantat-Gaudin2022}. In the present study, we select \gaia DR3 sources that are matched with the VIRAC2 data (see next section) in coordinates.

\subsection{VIRAC2}

The VVV survey \citep{Minniti2010} and its extension, the VVV eXtended survey (VVVX), acquired multi-epoch observations between 2010 to 2023 (hereafter we always refer to the survey as VVV, but it includes also the VVVX data). VVV produced a map of the Galactic bulge and southern part of the disc ($-130\deg<l<20\deg$ and $-15\deg<b<10\deg$) covering $\sim1700\deg^2$ in three near-infrared passbands: $J(1.25~\mu m)$, $H(1.64~\mu m)$ and $K_s(2.14~\mu m)$ (the VVV original footprint is also cover in $Z(0.87~\mu m)$ and $Y(1.02~\mu m)$ bands).
For further information about the VVV survey and its data quality, we refer to publications by \citet{Minniti2010}, \citet{Saito2012}, \citet{AlonsoGarcia2018}, and \citet{Surot2019a}.
VIRAC2 (Smith et al., in prep.) is the second data release of the VVV Infrared Astrometric Catalogue (VIRAC) \citep{Smith2018}. VIRAC2 is constructed from the PSF photometry catalogue, while VIRAC data are derived from aperture photometry. VIRAC2 is 90\% complete up to $K_s\sim16$ across the VVV bulge area \citep{Sanders2022}. Its proper motions are anchored to \gaia DR3 absolute reference frame. \rev{\citet{luna2023} tested the reliability of  VIRAC2 proper motions and their errors.}

In our study we use preliminary VIRAC2 data, and adopted the following quality selection criteria: 
(i) sources with a complete (5-parameter) astrometric solution, (ii) non-duplicates, (iii) sources that are detected in at least 20\% of the observations,\footnote{There are between $\sim 100-300$ $K_s$-band epochs per tile in the Milky Way central regions of the VVV.}, and (iv) sources with unit weight error (\texttt{uwe}) $uwe<1.2$. The latter is used as a threshold to select single sources with good astrometric measurements.

\section{Selection of Red Clump Sources}

Red clump (RC) stars are in the helium core-burning phase on their second ascent towards the giant branch \citep[see][for a review]{Girardi2016}. They are easily identified in nearby stellar systems because they occupy a well-defined region in Colour Magnitude Diagrams (CMDs). They have a characteristic luminosity and exhibit minimal variability, which makes them a useful standard candle for distance determination \citep[][]{Alves2002, RuizDern2018}.

Using the VIRAC2 data set, we select a region in the inner part of the Galactic bulge, in a box within $-5\deg<l<5\deg$ and $-3\deg<b<3\deg$. The total number of sources from VIRAC2 within the 60 sq.deg region is $112\,599\,161$.
In the same area, there are $36\,101\,569$ sources in \gaia.

\begin{figure}
    \centering
    \includegraphics[width=0.49\textwidth]{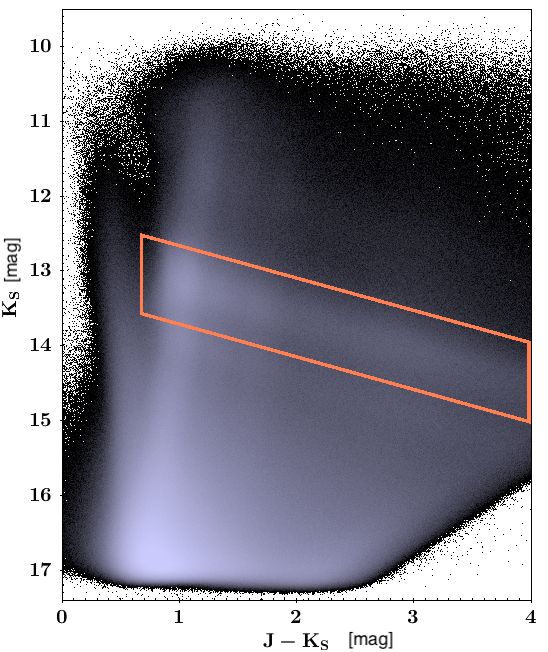}
    \caption{VVV near-IR CMD of the 60 sq.deg studied region towards the Galactic bulge. The orange box encloses the area used for the RC sources selection.}
    \label{fig:CMD_RC}
\end{figure}

In Figure \ref{fig:CMD_RC} we show the $J-K_s$ vs. $K_s$ CMD for all the stars in the 60 sq.deg bulge area centred on the GC. The blue sequence at $J-K_s \leq 0.6$~mag, populated by the Main Sequence disc, and the redder Red Giant Branch (RGB) bulge sequence merge at $K_s > 15$~mag. The RGB is the nearly vertical sequence ranging in colour around $J-K_s \sim 0.8-1.1$. Due to very high reddening in the centre-most regions of the bulge the RGB fans along the reddening vector covering the entire $J-K_s$ range of the CMD.
There are three overdensities along the Red Giant Branch (RGB) in the $J-K_s$ CMD (Fig. \ref{fig:CMD_RC}). While the most obvious one, which we highlight inside an orange box in Fig. \ref{fig:CMD_RC}, corresponds to the RC stars from the bulge, the nature of the other two overdensities is more uncertain \citep{AlonsoGarcia2018,Nataf2011,Gonzalez2018,Nataf2013}. 

In the CMD, we select the overdensity traced by the RC stars inside the orange box marked with a solid line in Figure \ref{fig:CMD_RC}. These stars are selected within $\pm0.5$mag along the reddening vector:

\begin{equation}
    K_s=12.7-0.428(J-K_s)
\label{eq:orangebox}
\end{equation}

We have used the $K_s$ total-to-selective ratio from \citep{AlonsoGarcia2017}:

\begin{equation}
    \frac{A_{K_s}}{E(J-K_s)}=0.428 \pm0.005 \pm0.04,
\label{eq:red_vect}
\end{equation}

\noindent where the infrared colour excess of the RC stars is $ E(J-K_s)=(J-K_s)-(J-K_s)_0 $ and each star has an individual value taken from the reddening map of \citet{Sanders2022}.

\begin{figure}
    \centering
    \includegraphics[width=0.49\textwidth]{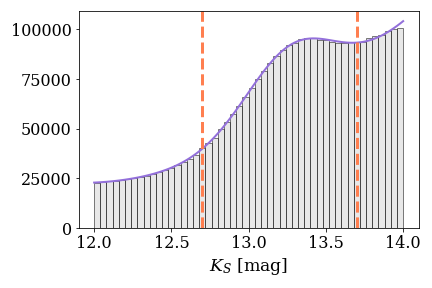}
    \caption{$K_s$ luminosity function around the less reddened RC location of the RGB selected around $(J-K_s)=1$. The solid line is fit to the distribution and follows Equation \ref{eq:LF}, and the dashed lines are the adopted limits of the RC selection.}
    \label{fig:Q1_LF}
\end{figure}

The orange box is centred on the RC. The intrinsic colour of the Galactic bulge RC is $(J-K_s)_0=0.68$ mag \citep{Alves2002,Gonzalez2012,RuizDern2018}, but the bulk of the RC in the bulge is at $J-K_s \sim 1.0$~mag. The brightness of the RC can be determined from the $K_s$ luminosity function (LF) shown in Figure \ref{fig:Q1_LF}. The LF around the bulge RC is fitted with a Gaussian plus a polynomial \citep{Gonzalez2013}: 

\begin{equation}
    N(K_s)=a+bK_s+cK_s^2+\frac{N_{RC}}{\sigma_{RC}\sqrt{2\pi}}\exp (-\frac{(K_s^{RC}-K_s)^2}{2\sigma_{RC}^2})
\label{eq:LF}
\end{equation}

The fitted parameters are: $a=-4.33\times 10^5,~b=1.82\times 10^4,~c=2.6\times 10^6,~\sigma_{RC}=0.3 mag,~K_s^{RC}=13.27$ mag.
\rev{; where $\sigma_{RC}=0.3 mag$ results from the convolution of the intrinsic width of RC due to the distribution of stellar masses that populate the RC in the observed population \citep[0.2\,mag,][]{Alves2000}, the observed width of the RC \citep[0.22\,mag,][]{Surot2019}, and the photometric error from VIRAC2 ($\sim$0.12\,mag around the RC)}. In Figure \ref{fig:Q1_LF}, the solid line is fit to the $K_s$ LF following Equation \ref{eq:LF}, and the dashed lines mark the limit of the RC selection.
The LF for stars around $(J-K_s)=1$ peaks at $K_s=13.3$ mag. To exclude foreground Main Sequence disc stars, we limit the RC selection polygon at the blue end to $(J-K_s) \geq 0.8$~mag.

\begin{figure}
    \centering
    \includegraphics[width=0.49\textwidth]{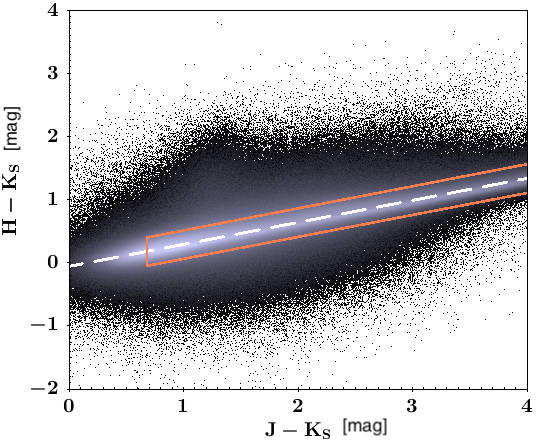}
    \caption{Colour-colour plot covering the 60 sq.deg area search box ($-5\deg<l<5\deg$ and $-3\deg<b<3\deg$). The dashed line is the linear fit to refine the RC selection. The orange box encloses further selection of RC stars.}
    \label{fig:Q1_CC}
\end{figure}

We then refine the selection of the RC candidates by a linear fit to the $H-K_s$ vs $J-K_s$ relation of the stars within the search box ($-5\deg<l<5\deg$ and $-3\deg<b<3\deg$), and selecting $\pm0.25$mag to the sides of the fit. Figure \ref{fig:Q1_CC} shows the $H-K_s$ vs $J-K_s$ diagram, the polygon marks the boundary of the RC selection and the dashed line is the linear fit:

\begin{equation}
    H-K_s=0.35(J-K_s)-0.06
\end{equation}

After selecting the RC region in the CMD and in the colour-colour diagram from VIRAC2, we crossmatch the sources with \gaia DR3 with a $0\farcs5$ tolerance in position, followed by removing spurious matches, which are identified as outliers in the distributions of $(G-K_s)$ colours.

The mean separation of the crossmatched sources is 0.03 arcsec.


The \gaia DR3 -- VIRAC2 matched sample consists of $2\,594\,052$ RC candidate stars within the 60 sq.deg region in the inner bulge.














\subsection{Foreground decontamination}

The principal contaminant of the sample can be  M-dwarf stars and reddened main sequence disc stars and red giant stars from the bulge that populate the same area as the RC stars in the CMD and colour-colour diagrams \citep{Mejias2022,AlonsoGarcia2018}.
\rev{Those giant stars, at the faint end of the RC LF, belong to the RGB bump \citep{Nataf2011}, which overlaps with the more distant RC giants tracing the Milky Way structure behind the galactic bar \citep{Gonzalez2018}. \citet{2021A&A...647A..34L} studied the RC population in the southern bulge ($-6\,$deg$<b<-9\,$deg) and found around 25\% of RGB bump interlopers in the magnitude range $13\,$mag$<K_{s_0}<14\,$mag. A spectroscopy follow-up can discern the two populations \citep[e.g.,][]{2022MNRAS.512.1710H}.}

Bright foreground stars have well-characterised, reliable parallax in \gaia DR3 and/or in VIRAC2. Hence we implement a cut in parallax removing those stars that lie within 5 kpc ($\varpi>0.2$ mas) and are bright enough to have precise parallax ($\frac{\sigma_{\varpi}}{\varpi}<0.5$). However, our sample might still contain faint foreground stars with poorly determined or no parallax measurements in the catalogues. We discuss this further in Section \ref{sec:selection of stars}.

After the parallax cut our sample was reduced from $2\,594\,052$  
to $2\,077\,026$ sources. 

\subsection{Population correction of RC absolute magnitude and distance determination}

The absolute magnitude of RC depends on metallicity and age.
The brightness of the RC in the Solar Neighbourhood is
$M_{K_s}=-1.61$ \citep{RuizDern2018,Sanders2022,Alves2002}.
To account for the metallicity and age dependence, a population correction is needed.

The metallicity distribution function (MDF) of RC stars in the bulge is bimodal \citep{Zoccali2017,2020MNRAS.499.1037R,2021A&A...656A.156Q, 2020A&A...638A..76Q}. The peaks are at around $[Fe/H]=-0.4$ and $[Fe/H]=+0.3$, with the contributions assumed to be 50\% each\footnote{The fractional contribution of metal-rich vs metal-poor population depends on Galactic latitude \citep[e.g.][]{Zoccali2017}, but this variation results in distance modulus difference of only $\sim0.005$ mag.}. We take from Table 1 of \citet{SalarisGirardi2002} the stellar evolutionary model theoretical RC magnitudes for metallicity $Z=0.008$ and $Z=0.03$, corresponding to the two peaks, and average the contributions of these two bulge components to get the mean absolute $K_s$ magnitude $<M_{K_s}>$.




We also consider three different ages: 8, 10 and 12 Gyr, which are typical of bulge stars \citep[e.g.,][]{Sit2020,Hasselquist2020,Surot2019}. For each age, we take the average $<M_{K_{s_{Z1,Z2}}}>$ given the two metallicity components.

The difference between the local theoretical RC absolute magnitude -- i.e., the absolute magnitude of the RC in the stellar evolutionary models of \citet{SalarisGirardi2002} --- and the mean absolute RC magnitude for a given age and metallicity mix is the so-called population correction. It is derived from the \citet{SalarisGirardi2002} models in the following way:

\begin{equation}
    \Delta M_{K_s}^{RC}= M_{K_s, theo}^{RC} - <M_{K_{s_{Z1,Z2}}}>
\end{equation}

\noindent where $ M_{K_s, theo}^{RC}=-1.54$.

The population correction ($\Delta M_{K_s}^{RC}$) variation between -0.011 and -0.099 mag, depending on age of the bulge RC stars, (see Table \ref{tab:pop corr}) is taken as the systematic error in the distance derivation. 

The distance modulus $\mu_0$ is then:

\begin{equation}
   \mu_0= m_{K_s}^{RC}- M_{K_s, obs.}^{RC} - A_{m_{K_s}} + \Delta M_{K_s}^{RC}
\end{equation}

\noindent where $M_{K_s, obs.}^{RC}$ is the observationally determined 
absolute $K_s$ magnitude for the local (Solar vicinity) RC population.

\begin{table}
    \centering
    \begin{tabular}{c|cc|c}
    \hline
    \hline
        Age (Gyr) & $Z=0.008$ & $Z=0.03$ & $\Delta M_{K_s}^{RC}$   \\\\
        \hline
         8 & -1.446 & -1.611 & -0.011 \\
         10 & -1.385 & -1.571 & -0.062 \\
         12 & -1.309 & -1.572 & -0.099 \\
    \hline         
    \end{tabular}
    \caption{Population correction ($\Delta M_{K_s}^{RC}$) for the $K_s$ absolute magnitude of RC stars. $\Delta M_{K_s}^{RC}$ is computed with Table 1 of \citet{SalarisGirardi2002} assuming different ages and metallicities of Galactic bulge stars.}
    \label{tab:pop corr}
\end{table}

To compute the distance, we assume $M_{K_s, obs.}^{RC}=-1.61\pm0.07$ from \citet{Sanders2022}, which is in agreement with \citet{RuizDern2018} $M_{K_s}=(-1.606\pm0.009)$.
Then, applying the population correction described above, the distance in pc is expressed as:

\begin{equation}
    d=10^{0.2(\mu_0+5)}
\end{equation}

The 
heliocentric distances for our selection of RC candidate stars in the bulge CMD range between 7 and 11.5 kpc. For a typical RC star in our sample ($K_s=13.2$ mag and extinction of $A_{K_s}$=0.26 mag), the population correction implies a difference in distance of ~130 pc compared to the distance derived only using the observed local absolute magnitude ($M_{K_s}=-1.54$ mag instead of $M_{K_s}-1.61$ mag).

Adding to that, by assuming different ages, that is, different corrections, the resulting distances vary by ~250 pc, which at the distance of the bulge would be a relative error of 3\%. Such value is added to the error budget of the distance distribution.

\subsection{Tangential velocity derivation}

The high and variable dust extinction towards the Galactic bulge, added to the crowded stellar environment, might cause \gaia DR3 proper motions not be well characterised \citep[][]{luna2023,Lindegren2018,Battaglia2022,Fabricius2021,Riello2021}.
For those reasons, we make a separate selection based only on VIRAC2 proper motions, derived from NIR observations, for the following analysis.

After removing the bright foreground objects and having the distance estimation for the RC sources, the VIRAC2 proper motions ($\mu_{\alpha^*}=\mu_{\alpha} \cos(\delta)$ and $\mu_{\delta}$) are corrected for the reflex motion of the Sun ($(U_{\odot}, V_{\odot}, W_{\odot})=(12.9, 245.6, 7.78)$~km~s$^{-1}$) 
using the {\tt gala} package \citep{Price2017} from {\tt Astropy} \citep{Astropy2013}. 
For the correction computation we adopt the default Solar motion relative to the GC as a combination for the peculiar velocity \citep{Sch2010} and for the circular velocity at the Solar radius \citep{Bovy2015}.
The tangential velocity of each star in the sample is:

\begin{equation}
    V_{tan}=4.74\, \mathrm{km\,s}^{-1} \times \left( \frac{\mu}{\mathrm{mas\,yr}^{-1}} \right) \times \left( \frac{d}{\mathrm{kpc}} \right)
\label{eq:vtan}
\end{equation}

\noindent where $\mu$ is the VIRAC2 proper motion and $d$ is the heliocentric distance.




\subsubsection{Monte Carlo sampling}
We use a Monte Carlo sampling approach to derive a tangential velocity distribution for each star in our final sample. The VIRAC2 catalogue provides the correlation coefficient between the RA and Dec components of the proper motions. Together with the proper motion uncertainties, one can use the covariance matrix of each source to generate a multivariate Gaussian distribution of the proper motion errors.

With that, we draw 100,000 random samples of proper motions for each star, which follow a Gaussian distribution. Each of those proper motions is corrected for the reflex motion of the Sun as described in the previous section. 

For the distance, we assume a Gaussian distribution with a width given by the following parameters added in quadrature:\rev{the photometric error from VIRAC2 ($\sim$0.12\,mag around the RC), the absolute magnitude calibration error \citep[0.07\,mag,][]{Sanders2022} and the population correction \citep[0.088\,mag,][]{SalarisGirardi2002}, resulting in a typical distance error of $\sim450\,$pc.}


From such distribution, we draw 100,000 random samples of distances for each source. Finally, for each source, we obtain 100,000 samples of tangential velocity, with the median of the distribution being their characteristic tangential velocity and the uncertainties are computed from the 16$\mathrm{^{th}}$ and 84$\mathrm{^{th}}$ percentile.

\section{Results and discussion}

The Gaia DR3-VIRAC2 crossmatch and foreground decontamination of the sample resulted in 2\,077\,026 sources located in the RC region of the NIR-CMD. 
We further restrict the selection of RC candidate stars to a narrower location in the CMD, within $0.3<(J-K_s)_0<1$ and $-0.1<(H-K_s)_0<0.4$.
From those, \rev{49} sources have a lower limit in tangential velocity (16$^\mathrm{th}$ percentile) above \rev{the local escape velocity}, with a relative error in their tangential velocities smaller than 30\% ($\frac{\sigma_{V_{tan}}}{V_{tan}}<0.3$) and $\sigma_{V_{tan}}$ the difference between the 16$^\mathrm{th}$ and 84$^\mathrm{th}$ percentiles.
The local escape velocity depends on the assumed potential, and for the Galactic bulge, it ranges between $\sim600$ and $\sim800$ \kms, reaching the latter within $1\,$kpc of the GC \citep{Rossi2017,Deason2019}. 

\rev{To be more inclusive, we adopt the lower limit in velocity as the escape velocity given by the \texttt{MWPotential2014} Galactic potential from the \texttt{galpy} package \citep{Bovy2015}, with the addition of the influence of the bar --using the \texttt{DehnenBarPotential} from \texttt{galpy} -- and a Kepler potential for SgrA*. The potential is in the low-mass end of the Galaxy mass estimates \citep{2019MNRAS.484.5453C}, with a total mass within 60\,kpc of $4\pm0.7\times10^{11}\,M_{\odot}$ \citep{2008ApJ...684.1143X}. In this potential, the dark-matter halo is described by a NFW Potential, the bulge is modelled as a power-law density profile with an exponent of $-$1.8 and a cut-off radius of 1.9\,kpc and the disc is modelled with a Miyamoto Nagai potential. 
The selection of the potential implies that some of the selected candidates move fast, but might still be bound, depending on their exact locations.}

Their tangential velocities range between \rev{731 and 1938} \kms, and are located between \rev{0.3 and 3.5 kpc} from the GC (considered to be at $R_0=8.122\,$kpc \citep{GRAVITY2018}). The range of tangential velocities is in agreement with the predictions from \citet{Generozov2020} and data from \citet{Koposov2020,Brown2015,Boubert2018}.

\subsection{Possible sample contaminants and selection of interesting objects}
\label{sec:selection of stars}

The current data do not have sufficient information to assign a spectral type to each star. A precise spectroscopic determination of the stellar parameters ($\log g$ and $T_{\mathrm{eff}}$) and the metallicity will enable us to confirm the stars as RC giants, hence confirming their location in the Galactic bulge.

To refine the sample of RC stars, we identify sources that might be \rev{M-dwarfs} given their colours. For this, we follow the colour cuts by \citet{Mejias2022}, based on a spectroscopically confirmed sample of M dwarfs by \citet{West2011}:

\begin{equation}
\begin{split}
   0.414<J-H<0.695 \\
   0.058 < H-K_s < 0.504 \\
   0.621 < J-K_s < 1.051
\end{split}
\label{eq:bds}
\end{equation}

\begin{figure}
    \centering
    \includegraphics[width=0.4\textwidth]{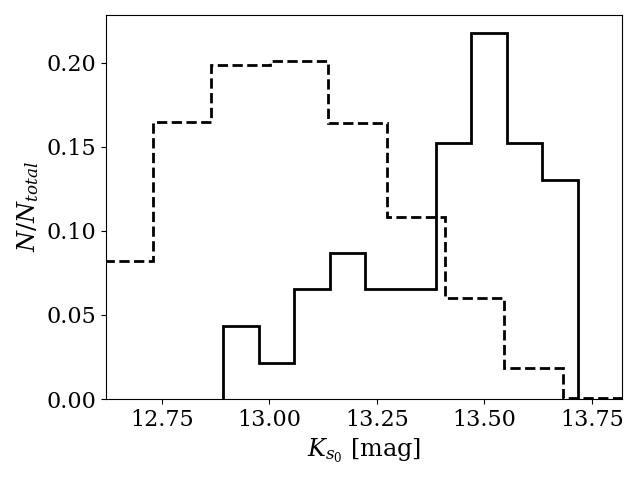}
    \caption{$K_{s_0}$ luminosity function. The y-axis shows the normalised count of sources. The dashed histogram are the stars within the magnitude range $12.5<K_{s_0}<13.75$. The solid histogram are the \rev{46} stars that travel with a tangential velocity larger than the local escape velocity and that are not identified as M-dwarfs by their colour.}
    \label{fig:LF_spcfoll}
\end{figure}

\rev{Three} sources fall into the selection cuts, placing them as possible \rev{M-dwarfs}. Their properties are listed in Appendix \ref{sec:appendixBDS_V2GDR3}.
Removing the likely \rev{M-dwarfs}, the selection results in a sample  of \rev{46} stars.
The excess of stars in the faint end of the LF (solid histogram in Fig. \ref{fig:LF_spcfoll}) is expected.
It is a consequence of a larger volume for the more distant, thus fainter, sources in the sample.

\begin{figure}
    \centering
    \includegraphics[width=0.49\textwidth]{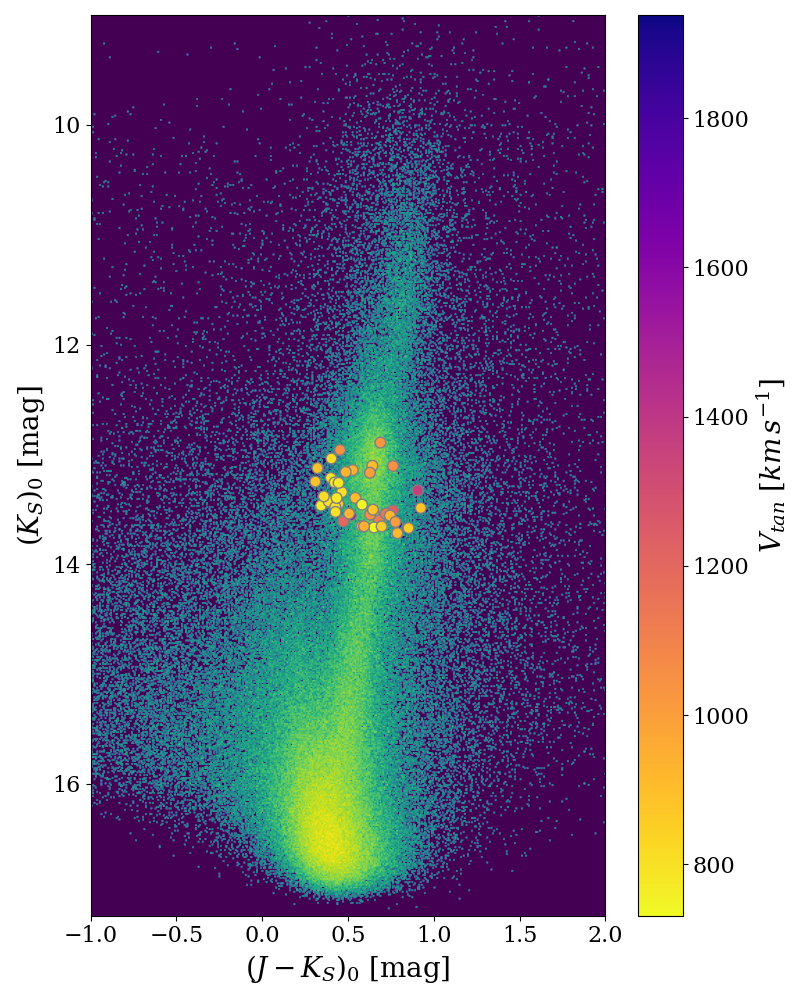}
    \caption{\rev{Deredenned CMD}. The colored points are stars that travel with a tangential velocity larger than the local escape velocity. }
    \label{fig:cmd_dered_hivtan}
\end{figure}

Figure \ref{fig:cmd_dered_hivtan} shows the location of the \rev{46} high-velocity stars in the extinction-corrected NIR CMD. 

Given the heliocentric distance $d$ derived from the RC calibration, we derive the distance with respect to the GC ($r_{GC}$)
using {\tt Astropy} and the following parameters:
the ICRS coordinates of the GC (RA,Dec)$=(266.4051, -28.936175)$\,deg \citep{Reid2004}; the velocity of the Sun in Galactocentric Cartesian coordinates $(U_{\odot},V_{\odot},W_{\odot})=(12.9, 245.6, 7.78)$\,\kms \citep{Drimmel2018,GRAVITY2018,Reid2004}; the distance from the Sun to the Galactic midplane $z_{\odot}=20.8$\,pc \citep{Bennett2019}, and the distance from the Sun to the GC $d_{\odot}=8.122$\,kpc \citep{GRAVITY2018}.

\begin{figure*}
    \centering
    \includegraphics[width=0.7\textwidth]{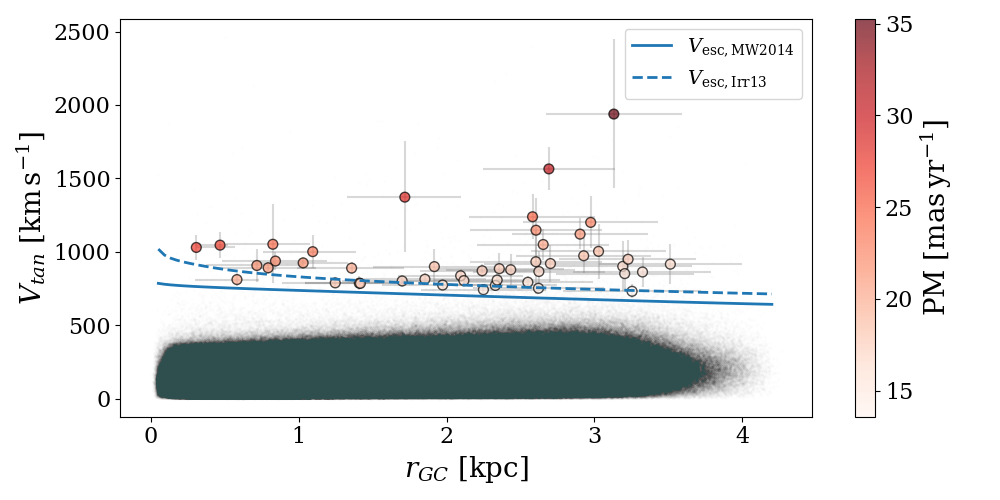}
    \caption{\rev{Tangential velocity vs galactocentric distance ($r_{GC}$)}. \rev{The solid and dashed lines are the escape velocity at a given $r_{GC}$ assuming the  \texttt{galpy} Galactic potential \texttt{MWPotential2014}, or \texttt{Irrgang13III}, respectively.}. The coloured points are the \rev{265} stars remaining after foreground objects' decontamination, and colour and quality cuts. The points are colour-coded by their VIRAC2 proper motions (PM).}
    \label{fig:vtan_rgc}
\end{figure*}

Figure \ref{fig:vtan_rgc} shows the tangential velocity of the stars as a function of the galactocentric distance ($r_{GC}$). \rev{The solid line indicates the escape velocity at a given galactocentric distance, assuming the \texttt{MWPotential2014} potential from the \texttt{galpy} package \citep{Bovy2015}. The adopted potential is only a reference that helped us to select the threshold in velocity.
Different assumptions would change the escape velocity by tens of \kms. As a reference for a higher escape velocity profile, the plot also shows the escape velocity profile given by the \texttt{Irrgang13III} potential from the \texttt{galpy} package. This potential is the Model III of \citet{2013A&A...549A.137I}, with a total Galaxy mass within 50\,kpc of $8.1\times10^{11}\,M_{\odot}$.} 

\begin{figure*}
    \centering
    \includegraphics[width=0.7\textwidth]{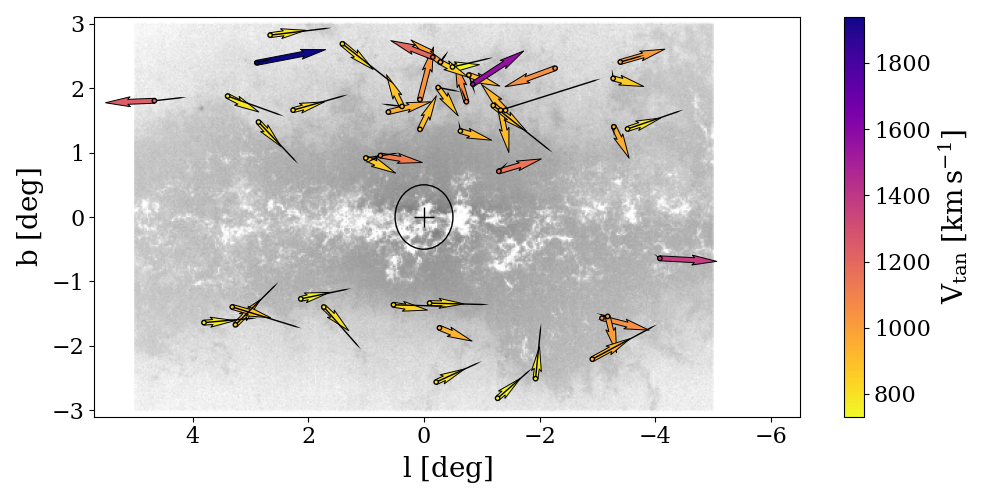}
    \caption{Stars that travel with a tangential velocity larger than the local escape velocity. The arrows represent the VIRAC2 proper motion vectors along the galactic latitude ($b$) and longitude ($l$) components after correcting for the reflex motion of the Sun. The points and arrows are colour-coded by the tangential velocity of the source. Black lines indicate the \gaia DR3 proper motion vector transformed to galactic coordinates and corrected for the reflex motion of the Sun. The cross places the GC at $(l,b)=(0,0)$ and the circle surrounding it represents a radius of $0.5\deg\simeq75\,$pc.}
    \label{fig:vpd_hivtan}
\end{figure*}

The position in the sky of the final sample is shown in Fig. \ref{fig:vpd_hivtan} in galactic coordinates. The background is the density plot of the VIRAC2 sources in the search box, where a whiter colour indicates a higher density, and is  
equivalent 
to a reddening map. The arrows represent the VIRAC2 proper motion vectors along the galactic latitude ($b$) and longitude ($l$) components, colour-coded by their tangential velocity.
Black lines indicate the \gaia DR3 proper motion vector transformed to galactic coordinates and corrected for the reflex motion of the Sun. We note that the \gaia DR3 proper motion of some sources matches with that of VIRAC2, particularly for the sources with slower velocities.

In Fig. \ref{fig:vpd_hivtan}, we note asymmetries in the spatial distribution and proper motions of the sample. In the initial sample of \gaia DR3 -- VIRAC2 crossmatch, the stars are distributed evenly in galactic latitude and longitude. However, in the sample of \rev{46 high-velocity stars,  31 (67\%)} stars are located at positive galactic latitudes. Furthermore, \rev{39 (85\%)} stars have VIRAC2 proper motion vectors pointing towards negative galactic longitude.

\begin{figure}
    \centering
    \includegraphics[width=0.49\textwidth]{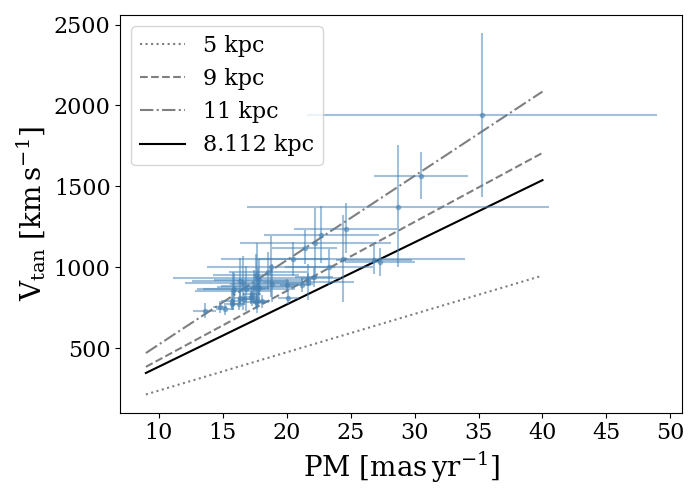}
    \caption{Tangential velocity as a function of proper motion at different fixed heliocentric distances. The blue points are the \rev{46} high-velocity stars in the sample described in Sect. \ref{sec:selection of stars}, their proper motions are from VIRAC2 and the tangential velocities are computed with Eq. \ref{eq:vtan}. The dashed lines indicate the tangential velocity as a function of the proper motion for different fixed heliocentric distances, while the solid line is at the distance of the GC.}
    \label{fig:pm_Vtan_dist}
\end{figure}

Based on the Gaia DR3 mock catalogue \citep{2020PASP..132g4501R}, a synthetic Milky Way catalogue that simulates the \gaia EDR3\footnote{\gaia EDR3 refers to the early data release, which contains the same astrometric and photometric information as DR3.} content,
within the magnitude and colour range of our sample ($17.8<G<19.7$\,mag;\, $2.6<BP-RP<3.9$\,mag), it is expected that $\sim$80\% of stars lie at a distance between 6.5 and 9.5 kpc. However, some of the stars in our final sample might still be foreground contaminants. Figure \ref{fig:pm_Vtan_dist} shows the tangential velocity of the stars as a function of their VIRAC2 proper motions. The dashed lines indicate the tangential velocity as a function of the proper motion for different fixed heliocentric distances, while the solid line is at the distance of the GC (8.122 kpc). Even if some of the stars were located at smaller heliocentric distances than our estimates, more than half of them are interesting objects travelling faster than the average disc rotation (246 \kms).

\subsection{HVSs candidates and flight time}
\label{sec:HVSs V2_GDR3}

To test whether the HVS candidates radiate from the GC, we project the proper motion vectors, appropriately account for the Solar motion, and search for those that intersect a region within $0.5\deg$ ($\sim 75$ pc at the distance of the GC) centred on the GC. 
This criterion is based on \texttt{Speedystar}\footnote{\href{https://github.com/fraserevans/speedystar}{https://github.com/fraserevans/speedystar}} \citep{Evans2022}, where we simulate ejections of stars from the GC via the Hills mechanism
and analyse the position angle of their proper motions corrected by the reflex motion of the Sun. 
The stars that are within 4 kpc from the GC, such as the stars in our sample, have a 2D velocity vector (projected proper motion) that approaches the GC in less than 0.45 deg. The criterion also accounts for the uncertainty in the GC distance (31\,pc$\sim$0.2deg), the GC position (0.66\,mas) \citep{GRAVITY2018,2021A&A...647A..59G}, and encloses the nuclear star cluster (0.03 deg) \citep{2014A&A...566A..47S,2023A&A...672L...8S}.

\begin{figure*}
    \centering
    \includegraphics[width=\textwidth]{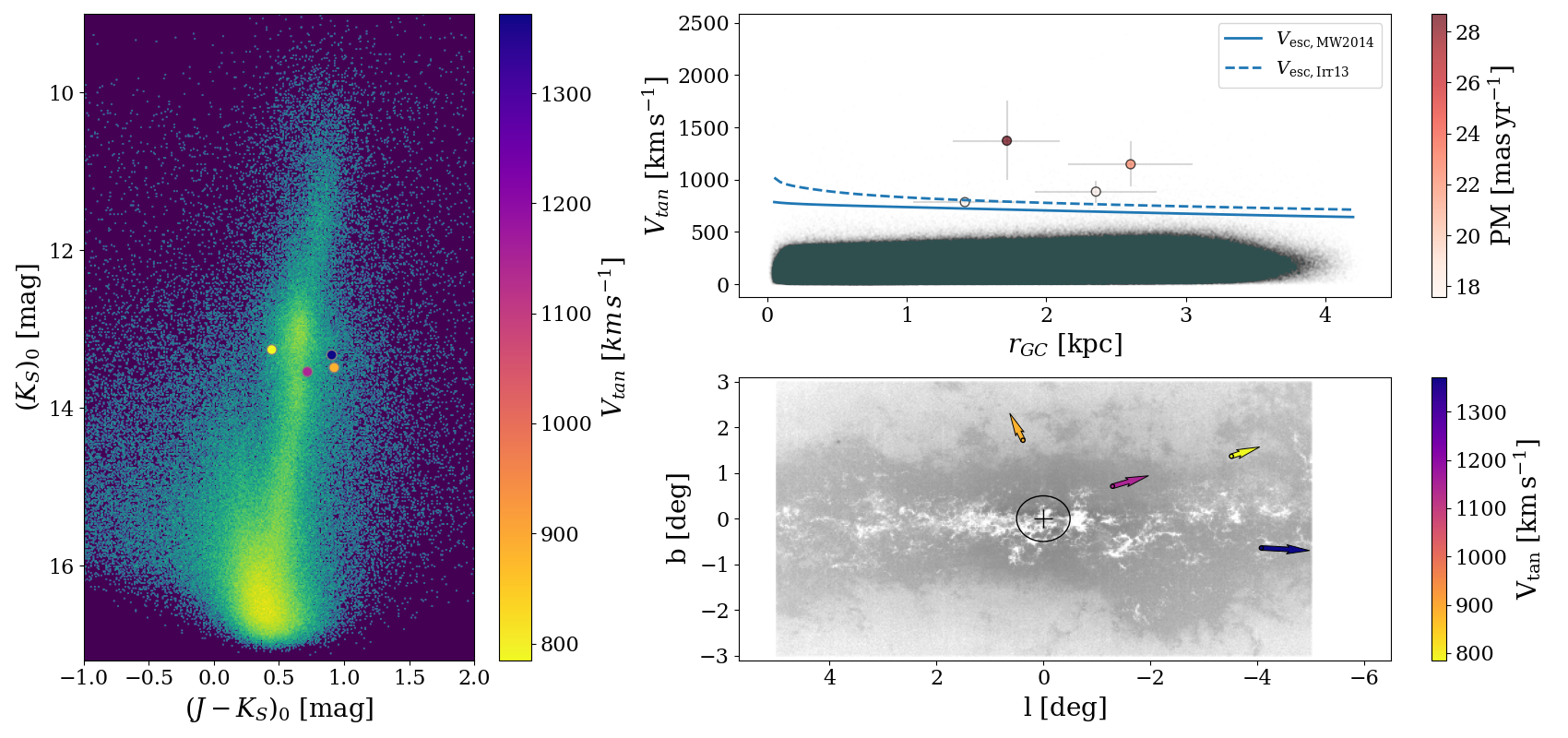}
    \caption{Near-infrared photometry, position and kinematics of the candidate HVSs detected in both VIRAC2 and \textit{Gaia} DR3. Stars that appear to be ejected from within $0.5\, \deg \simeq75$ pc of the GC. The left panel is the CMD, the same as in Fig. \ref{fig:cmd_dered_hivtan}. The top-right panel is the tangential velocity as a function of galactocentric distance, same as Fig.\ref{fig:vtan_rgc}. The bottom-right panel is the location of the stars in galactic coordinates with the VIRAC2 proper motions shown as arrows colour-coded by their tangential velocities, same as Fig. \ref{fig:vpd_hivtan}. In the bottom-right panel, the cross marks the GC, and the surrounding circle has a radius of $0.5\deg\simeq 75\,$pc.}
    \label{fig:from_ctr}
\end{figure*}

There are \rev{4} sources that fulfil this condition. Their projected 2D velocity vector approaches the GC in a range between $0.004\,\deg$ and $0.47\,\deg$; since these measurements lack radial velocity, the approach is not a physical quantity.
The left panel of Fig. \ref{fig:from_ctr} shows their location in the CMD colour-coded by their tangential velocities. The top-right panel shows their tangential velocity as a function of their galactocentric distance colour-coded by their VIRAC2 proper motions, while the bottom-right panel shows their location in galactic coordinates, with the VIRAC2 vectors represented as arrows colour-coded by the tangential velocity. 

To compute how long ago the HVS were ejected from the GC --the flight time, we project the VIRAC2 proper motion vectors backwards, in the direction towards the GC. We assume that the  proper motion direction is not strongly affected by the projection in the sky:

\begin{equation}
    t_f=\frac{\sqrt{l^2+b^2}}{\mu}\times3.6\times10^6 
\label{eq:flighttime}
\end{equation}

\noindent where $t_f$ is the flight time in years, $l$ and $b$ are in degrees and $\mu=\sqrt{\mu_{l^*}^2+\mu_b^2}$ in mas yr$^{-1}$.

\rev{The flight time of these 4 stars ranges from $1.1\times10^6$ yr to $3.2\times10^6$ yr.} We can roughly estimate the ejection rate of RC stars by dividing the number of stars over the range of flight times, this is $N_{HVS} / ( max(t_F) - min(t_F) )$, where $N_{HVS}$ is the observed number of Red Clump HVS candidates, and the denominator is the difference between the maximum and minimum flight time of the candidates. 
\rev{This corresponds to an ejection rate of $1.9\times10^{-6}$ yr$^{-1}$.} However, we note that this is not a complete sample, and it is limited to the RC mass range, typically between 0.8 and 2 $M_{\odot}$ \citep[][]{Girardi2016}. Our sample is also affected by contamination from non-RC stars that can only be disentangled with a spectroscopic follow-up. 
Therefore, the ejection rates are lower than the total integrated rate, which current studies estimate at $10^{-4}\,yr^{-1}$. This ejection rate estimate depends on the adopted IMF; if the IMF is top-light, the ejection rate can be $10^{-2}\,yr^{-1}$, but if it is top-heavy, it can be $10^{-4.5}\,yr^{-1}$ \citep[e.g.,][]{Evans2022,Rossi2017,Marchetti2022, 2022MNRAS.512.2350E}. 

Table \ref{tab:all} shows for the \rev{4} stars, the location in equatorial and galactic coordinates, and photometric information ($K_s$ and $J-K_s$), Table \ref{tab:all2} shows their VIRAC2 proper motions, and the derived parameters: tangential velocity ($V_{tan}$), galactocentric distance ($r_{GC}$), flight time ($t_f$), closest approach of the 2D velocity vector to the GC, \rev{and the probability ($P$) of exceeding the escape velocity, given by  $V_{tan, 16^{\rm th}}(r_{GC})/V_{esc}(r_{GC})$, with $V_{tan, 16^{\rm th}}$ the lower limit in tangential velocity (16$^{\rm th}$ percentile), and $V_{esc}(r_{GC})$ given by the \texttt{galpy} \texttt{Irrgang13III} potential; the stars that exceed such escape velocity have a probability of 1}. 

\begin{table*}
    \centering
    \begin{tabular}{lcccccc}
    \hline
    \gaia DR3 source ID & RA [hms] & Dec [dms] & l [deg] & b [deg] & $K_s$ [mag] & $J-K_s$ [mag]\\
        \hline
        \hline
4060857768397863040 & 17h39m53.68s & -27d42m35.30s & 0.3801 & 1.7167 & 13.82 & 1.69 \\
4057025592419323520 & 17h39m43.89s & -29d39m47.42s & -1.2939 & 0.7088 & 14.19 & 2.24 \\
4058312403271263360 & 17h31m35.90s & -31d10m44.87s & -3.5167 & 1.3669 & 13.62 & 1.3 \\
4054179029853634816 & 17h38m08.01s & -32d44m09.58s & -4.0754 & -0.6413 & 14.04 & 2.57 \\
        \hline
    \end{tabular}
    \caption{Location and photometric properties of the stars with \gaia DR3 counterpart and VIRAC2 proper motions pointing away from $0.5\deg$ around the GC. The Galactic longitude and latitude have a precision of $0\farcs5$.}
    \label{tab:all}
\end{table*}

\begin{table*}
    \centering
    \begin{tabular}{lccccccc}
    \hline
    \gaia DR3 source ID & $\mu$ [$mas\,yr^{-1}$] & d [kpc] & $r_{GC}$ [kpc] & $V_{tan}$ [\kms] & $t_f$ [yr] & Approach [deg] & P \\
        \hline
        \hline
4060857768397863040 & 17.77 & 10.46 & 2.36 & 885.15 & 1501668 & 0.379 & 1 \\
4057025592419323520$^*$ & 22.25 & 10.72 & 2.6 & 1147.16 & 1126034 & 0.297 & 1 \\
4058312403271263360$^*$ & 17.58 & 9.42 & 1.42 & 784.82 & 3239459 & 0.165 & 0.94 \\
4054179029853634816 & 28.69 & 9.72 & 1.72 & 1372.12 & 2772852 & 0.369 & 1 \\
        \hline
    \end{tabular}
    \caption{Derived parameters of the stars with \gaia DR3 counterpart and VIRAC2 proper motions pointing out from $0.5\deg$ around the GC. The second to last column indicates how close the projected 2D velocity vector approaches the GC. \rev{The last column indicates the probability of exceeding the escape velocity.}\\
    \rev{* According to the classifier of spurious astrometric solution in \textit{Gaia} DR3 \citep{2022MNRAS.510.2597R}, these sources have a good astrometric solution. The VIRAC2 proper motions are consistent within $1\sigma$.}}
    \label{tab:all2}
\end{table*}

\section{The red sample}

In this section, we explore those sources that appear in VIRAC2, but not in \textit{Gaia} DR3, because they are too faint or highly reddened.

We follow the procedure described in Sec. 3 and 4, to select the RC stars, and clean the sample of foreground stars including \rev{M-dwarfs} using VIRAC2 parallax and colour cuts. 

The initial sample consists of 8\,596\,826 sources that are in the VIRAC2 catalogue, but do not have a crossmatch in \textit{Gaia} DR3.
From those, we remove stars closer than 5 kpc to us ($\varpi>0.2$) and precise VIRAC2 parallax ($\frac{\sigma_{\varpi}}{\varpi}<0.5$). The remaining sample has 8\,214\,069 sources. \rev{We select sources with lower limits in tangential velocity }
\rev{larger than the escape velocity at a given galactocentric distance.} The resulting sample has \rev{504} sources from which \rev{481} have a relative error in their tangential velocities smaller than 30\%. From those sources, \rev{9} are possible \rev{M-dwarfs}, identified using the colour cuts of Eq.\ref{eq:bds}. Thus, the final sample has \rev{472} sources with tangential velocities up to \rev{$V_{tan}=2468$\,\kms.} These stars are shown in Fig. \ref{fig:V2notinGdr3}. The left panel shows the location of the sources in the CMD corrected for extinction. The top-right panel shows the tangential velocity as a function of the galactocentric distance ($r_{GC}$). The bottom-right panel shows the location of the sources in galactic coordinates, colour-coded by their tangential velocity, and with the proper motion vector shown as arrows.

\subsection{HVSs candidates and flight time}

\begin{figure*}
    \centering
    \includegraphics[width=\textwidth]{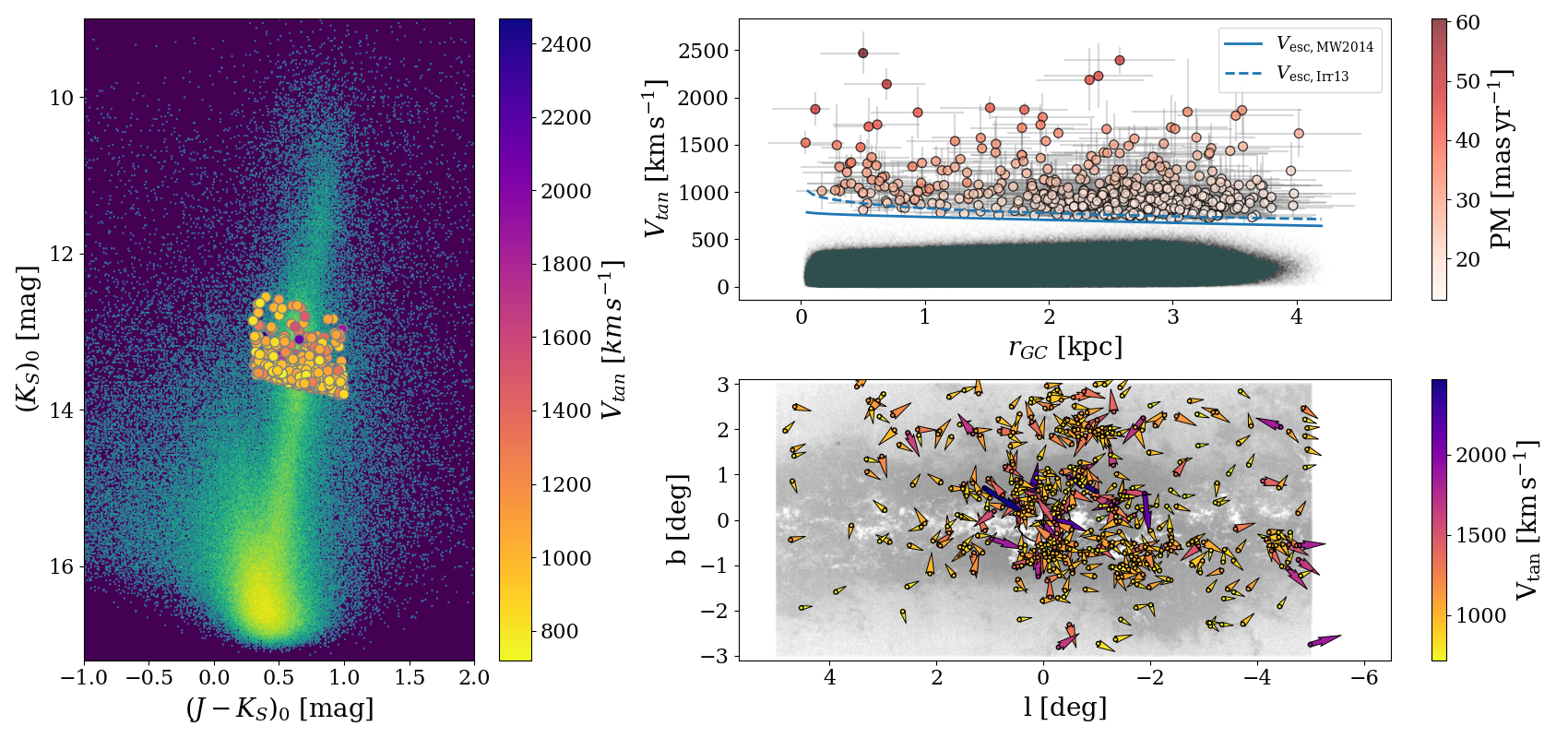}
    \caption{Near-infrared photometry, position and kinematics of the RC stars detected in VIRAC2, but not in \textit{Gaia} DR3. The left panel is the CMD, same as in Fig. \ref{fig:cmd_dered_hivtan}; the top-right panel is the tangential velocity as a function of galactocentric distance, same as Fig.\ref{fig:vtan_rgc}; and the bottom-right panel is the location of the stars in galactic coordinates with the VIRAC2 proper motions shown as arrows colour-coded by their tangential velocities, same as Fig. \ref{fig:vpd_hivtan}.}
    \label{fig:V2notinGdr3}
\end{figure*}

\begin{figure*}
    \centering
    \includegraphics[width=\textwidth]{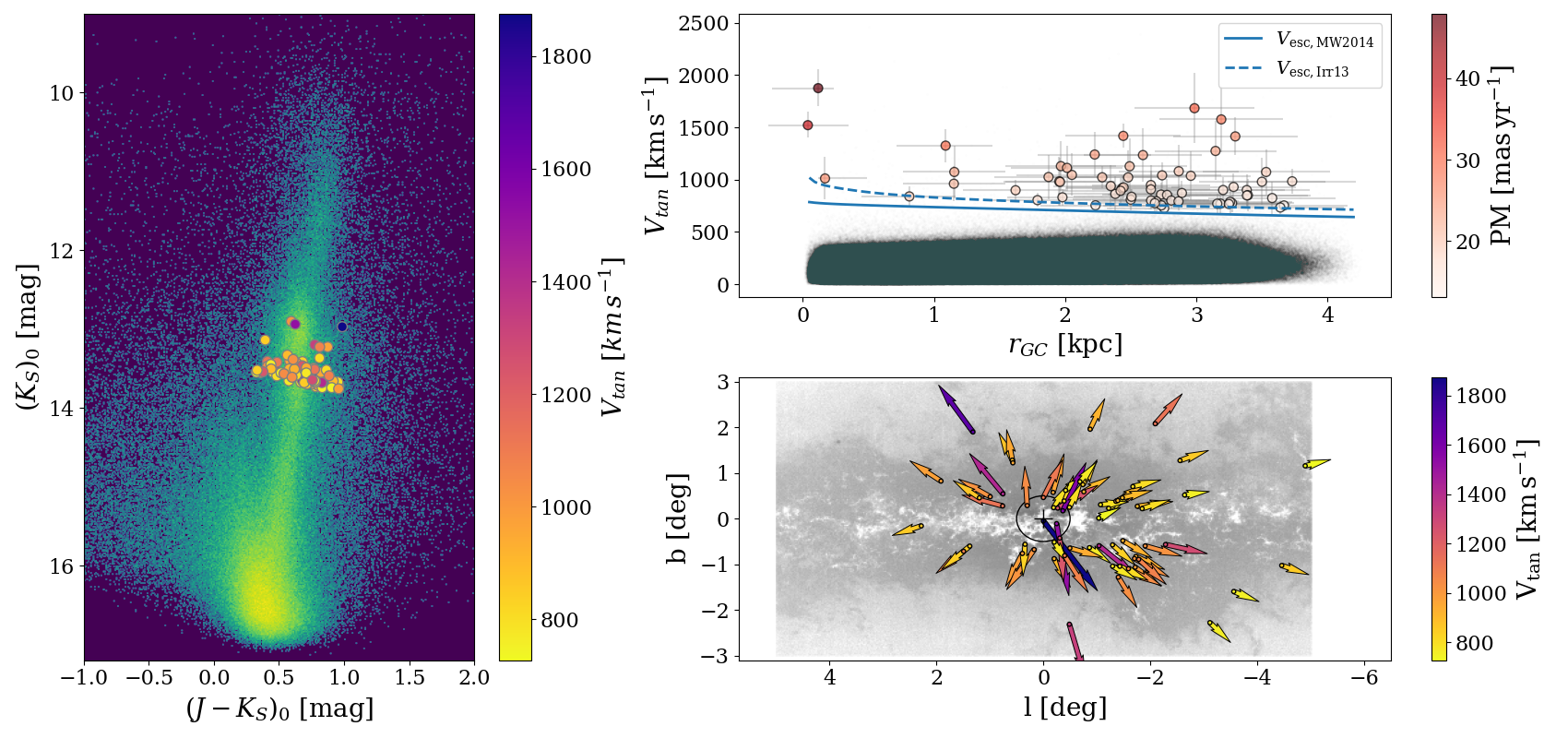}
    \caption{Near-infrared photometry, position and kinematics of the candidate HVSs detected in VIRAC2, but not in \textit{Gaia} DR3. Stars that appear to be ejected from within $0.5\,deg\simeq75\, pc$ of the GC. The left panel is the CMD, same as in Fig. \ref{fig:cmd_dered_hivtan}; the top-right panel is the tangential velocity as a function of galactocentric distance, same as Fig.\ref{fig:vtan_rgc}; and the bottom-right panel is the location of the stars in galactic coordinates with the VIRAC2 proper motions shown as arrows colour-coded by their tangential velocities, same as Fig. \ref{fig:vpd_hivtan}.}
    \label{fig:from_ctr_V2notinGdr3}
\end{figure*}

Figure \ref{fig:from_ctr_V2notinGdr3} is equivalent to Fig. \ref{fig:V2notinGdr3}, but shows the stars whose VIRAC2 proper motions point away from the GC.
We note that in contrast to Fig. \ref{fig:vpd_hivtan}, the density of sources is higher towards lower galactic latitudes, where \textit{Gaia} DR3 has fewer sources.
Similar to Fig. \ref{fig:vpd_hivtan}, \rev{72\% }of the sources have VIRAC2 proper motions pointing towards negative galactic longitude, but they are distributed evenly in latitude with respect to the Galactic plane.

There are \rev{65} sources that appear to be ejected from a region within $0.5\deg$ from the GC, with proper motion vectors pointing out of it and with tangential velocities up to \rev{$V_{tan}=1875$\,\kms.} 
Fig. \ref{fig:from_ctr_V2notinGdr3} shows the \rev{65} stars.

Their projected 2D velocity vector \rev{approaches the GC in a range between $0.0004\,\deg$ and $0.49\,\deg$}; as mentioned in Section 4.2, these values are not a physical quantity.

\begin{figure}
    \centering
    \includegraphics[width=0.45\textwidth]{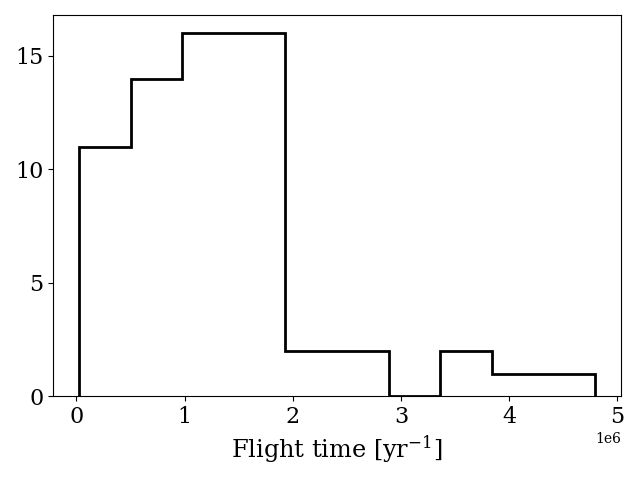}
    \caption{Flight time in Myr for the candidate HVSs that appear in VIRAC2, but not in \textit{Gaia} DR3 and have a proper motion vector pointing away from a location within $1 \,\deg\simeq75$ pc of the GC.}
    \label{fig:flighttime_V2notinGdr3}
\end{figure}

Following Eq.\ref{eq:flighttime}, the flight time of the \rev{65 stars appearing to be ejected from the centre ranges between $2.4\times10^4\,yr$ and $4.8\times10^6\,yr$.}
This provides a rough estimate of an ejection rate of \rev{$1.4\times10^{-5}\,yr^{-1}$,} an order of magnitude higher than for the sample that includes also \gaia DR3 measurements. As explained in Sec. \ref{sec:HVSs V2_GDR3}, this estimate is lower than the integrated ejection rate.
Figure \ref{fig:flighttime_V2notinGdr3} shows the distribution of flight times, with the major ejection episode occurring before 2\,Myr ago. 

Table \ref{tab:frmctr_v_extract} is an extract of Table \ref{tab:frmctr_v} and shows for the \rev{65} stars with VIRAC2 and no \gaia DR3 counterpart, the location in equatorial and galactic coordinates, and photometric information ($K_s$ and $J-K_s$). Table \ref{tab:frmctr_v_extract2} is an extract of Table \ref{tab:frmctr_v2} and shows VIRAC2 proper motions, and the derived parameters: tangential velocity ($V_{tan}$), galactocentric distance ($r_{GC}$), flight time ($t_f$), the closest approach of the 2D velocity vector to the GC, \rev{and the probability ($P$) of exceeding the escape velocity}.

HVSs can also originate from 
a globular cluster hosting an intermediate-mass black hole (IMBH).
Stellar interactions with the IMBH might result in the ejection of HVSs in a similar way as the Hills mechanism \citep[e.g.,][]{Fragione2019}. In that case, the star would point back to its parent cluster. 

The spatial and velocity distributions can be used to discern those stars from (hyper)runaway stars originating from the Galactic disc, where the runaway stars have lower velocities and are closer to the Galactic plane  \citep[e.g.,][]{Fragione2019,Generozov2022}. 
In the same way, the velocity distribution of HVSs originating from the GC would have a larger contribution to high-velocity stars.

A binary massive black hole (bMBH), like a pair of SMBH-IMBH, would produce HVSs with potentially anisotropic velocity distribution, depending on the eccentricity and mass ratio of the binary stellar system that gets disrupted \citep[e.g.,][]{1996NewA....1...35Q,2006MmSAI..77..653H,2006ApJ...651..392S,2006MNRAS.372..869D}. The velocity and spatial distribution predicted by the Hills mechanism are isotropic, thus, the anisotropy in the velocity and spatial distribution of HVSs originating from a bMBH can potentially distinguish their ejection scenario from the SMBH interaction.

\begin{figure}
    \centering
    \includegraphics[width=0.45\textwidth]{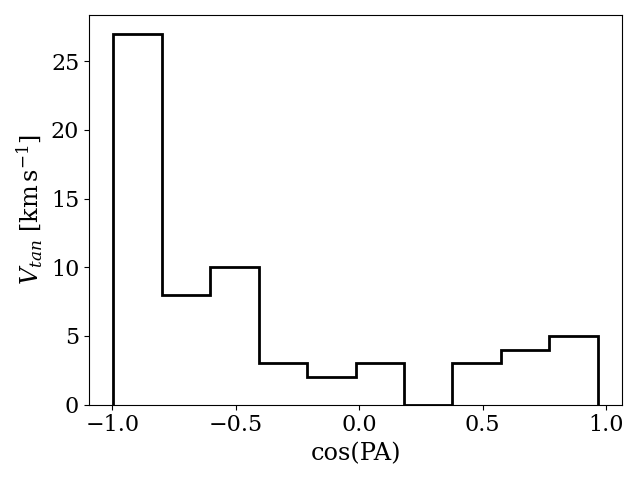}
    \caption{Cosine of the position angle of the stars in VIRAC2 that point out from $0.5\deg$ around the GC.}
    \label{fig:PA}
\end{figure}

\begin{figure}
    \centering
    \includegraphics[width=0.45\textwidth]{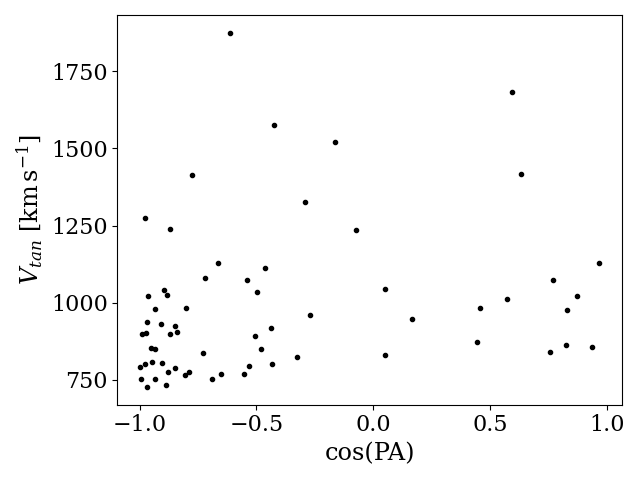}
    \caption{Tangential velocity as a function of the Cosine of the position angle of the stars in VIRAC2 that point out from $0.5\deg$ around the GC.}
    \label{fig:vtanPA}
\end{figure}

The spatial distribution predicted by the Hills mechanism is isotropic. In Fig. \ref{fig:PA}, we show the position angle (PA) of the sample of \rev{65} stars that point away from the GC. There is a peak in the $cos(PA)$ since most of the stars point towards negative longitudes, but without a trend with velocity (see Fig. \ref{fig:vtanPA}). Hence, the direction is not isotropic, but the velocity is.

\begin{figure}
    \centering
    \includegraphics[width=0.45\textwidth]{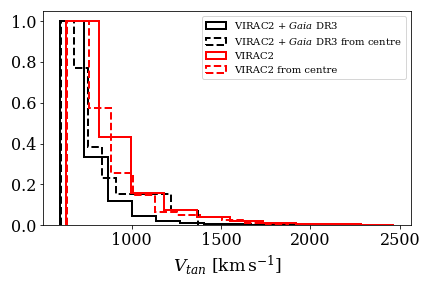}
    \caption{Inverse cumulative distribution of the high-velocity stars. The black histogram is for the final sample of stars in the VIRAC2 and \textit{Gaia} DR3 crossmatch (solid) and those that point out from $0.5\deg$ around the GC (dashed). The black histogram is for the final sample of stars in VIRAC2 (solid) and those that point out from $0.5\deg$ around the GC (dashed).}
    \label{fig:vtan_distr}
\end{figure}

Figure \ref{fig:vtan_distr} shows the inverse cumulative distribution function of the high-velocity stars' tangential velocity in both samples: 
black is the distribution of sources matched in VIRAC2 and \textit{Gaia} DR3 and that for sources without \textit{Gaia} DR3 counterpart is plotted with red line. The stars in the sample with only VIRAC2 data are closer to the GC. This sample also contains stars that reach higher velocities than the sample with \gaia DR3 counterpart. With the deceleration occurring in the first $\sim200\,$pc \citep{Kenyon2008}, the stars with \textit{Gaia} DR3 counterpart, that lie at larger distances from the GC, might have been decelerated.

\begin{table*}
    \centering
    \begin{tabular}{rcccccc}

\hline
ID & RA [hms] & Dec [dms] & l [deg] & b [deg] & $K_s$ [mag] & $J-K_s$ [mag] ]\\
\hline
\hline
1 & 17h51m54.16s & -28d01m38.73s & 1.4904 & -0.7146 & 14.28 & 2.26 \\
2 & 17h49m31.40s & -28d59m43.91s & 0.3911 & -0.7613 & 14.2 & 2.47 \\
3 & 17h48m40.41s & -29d08m22.46s & 0.172 & -0.6762 & 13.76 & 2.6 \\
4 & 17h51m11.57s & -28d04m06.82s & 1.375 & -0.6009 & 13.9 & 2.17 \\
5 & 17h51m33.79s & -27d03m47.14s & 2.2813 & -0.1578 & 14.83 & 3.73 \\
\multicolumn{7}{c}{\vdots}\\
\hline
\end{tabular}
\caption{Location and photometric properties of the stars with no \gaia DR3 counterpart and VIRAC2 proper motions pointing away from $0.5\deg$ around the GC. The Galactic longitude and latitude have a precision of $0\farcs5$. This is an extract, the complete table can be found in Appendix \ref{ap:table_virac2notGDR3_frmctr}.}
\label{tab:frmctr_v_extract}

\end{table*}

\begin{table*}
    \centering
    \begin{tabular}{rccccccc}

\hline

ID &  $\mu$ [$mas\,yr^{-1}$] & d [kpc] & $r_{GC}$ [kpc] & $V_{tan}$ [\kms] & $t_f$ [yr] & Approach [deg] & P \\
\hline
\hline
1 & 19.08 & 11.64 & 3.53 & 1072.9 & 1362189 & 0.344 & 1.0 \\
2 & 19.99 & 10.07 & 1.96 & 982.98 & 68908 & 0.001 & 0.97 \\
3 & 26.28 & 7.98 & 0.17 & 1011.28 & 489957 & 0.234 & 0.86 \\
4 & 19.82 & 8.9 & 0.81 & 839.35 & 1213281 & 0.383 & 0.89 \\
5 & 16.6 & 10.83 & 2.73 & 856.75 & 2020771 & 0.496 & 1.0 \\
\multicolumn{7}{c}{\vdots}\\
\hline
\end{tabular}
\caption{Derived parameters of the stars with no \gaia DR3 counterpart and VIRAC2 proper motions pointing out from $0.5\deg$ around the GC. The second to last column indicates how close the projected 2D velocity vector approaches the GC. \rev{The last column is the probability of exceeding the escape velocity}. This is an extract, the complete table can be found in Appendix \ref{ap:table_virac2notGDR3_frmctr}.}
\label{tab:frmctr_v_extract2}

\end{table*}

\section{Conclusions}

We have identified \rev{69} HVS candidates in the MW bulge that appear to be ejected from the GC. From those, \rev{4} have a \textit{Gaia} DR3 counterpart, but with \textit{Gaia} proper motions that are not reliable and \rev{65} were detected only in the VIRAC2 sample. All these stars are selected among the likely RC candidates that place them within the bulge. Their flight time ranges between \rev{$2.4 \times 10^4$ and $4.8 \times 10^6$~yr,} and implies an integrated \rev{ejection rate of the order of $1.4 \times 10^{-5}$ yr$^{-1}$,} provided that they all originated from the GC and are currently still within the bulge. Nevertheless, since the surveys are not complete, our sample is not complete. Adding to that, it is limited to RC stars, thus the derived ejection rates are lower than the current estimates ($10^{-4}\,yr^{-1}$). 

To fully determine the HVS candidates' spatial movement and reconstruct their orbits, we need to obtain the missing radial velocities.
The spectroscopic follow-up will also provide the chemical composition and spectral type measurements necessary to confirm the candidates as RC stars.
In the case a given star is not an RC star, the distance estimation would not be correct, nor its tangential velocity. However, such stars might be still interesting as they would be moving faster than the average disc rotation. 
Hence, they would offer the possibility of exploring the production scenarios of such stars through, for example, binary interactions or supernova kicks.

Having the first detection of HVSs in the Galactic bulge confirmed by future spectroscopy, we will be in a position to constrain the existing models, compare the observed with HVSs mock population \citep{Marchetti2018} and compute their orbits \citep[e.g.,][GravPot16\footnote{\href{https://gravpot.utinam.cnrs.fr}{https://gravpot.utinam.cnrs.fr}}]{FernandezTrincado2020}, and thus probe if they come from the GC or have another origin.

The Hills mechanism is parameterised by the shape of the stellar initial mass function in the GC, the distribution of binary orbital periods and mass fraction, and the ejection rate. 
The number of confirmed HVSs and their properties will constrain primarily the ejection rate and the shape of the IMF in the adopted model \citep[e.g.,][\texttt{Speedystar}]{Evans2022}. Interestingly, the discovery of the first HVSs close to the production centre will give a unique possibility to explore a recent stellar interaction with Sgr A*.


The velocity distribution is compatible with current ejection models, where the high-velocity stars should reside close to the GC, and those ejected with the Hills mechanism will also have an isotropic spatial distribution. A different spatial distribution, or proper motions that do not point back to the GC can be explained by other ejection scenarios, such as a disruption of globular clusters by the SMBH, inspiralling IMBH, dynamic interactions with binary massive BH, or supernova kicks. The high-velocity stars point back to the production origin, being these globular clusters, satellite galaxies such as the Magellanic Clouds, or the Galactic disc, where the runaway scenario is more probable. 

The future \textit{Gaia} data releases\footnote{\href{https://www.cosmos.esa.int/web/gaia/science-performance}{https://www.cosmos.esa.int/web/gaia/science-performance}} will improve the astrometry precision by a factor of $\sim2.5$, reaching $0.5$ mas yr$^{-1}$ and $0.3$ mas yr$^{-1}$ in DR4 and DR5, respectively, for the faintest stars ($G=20.7$ mag). The accuracy will improve as well, as the baseline of observations will increase to 60 months in DR4 and 120 months in DR5. The radial velocities are limited to sources brighter than $G\sim15.5$ mag and the spectra at $G\sim17.5$ mag. Thus, to confirm the nature of the candidate HVSs in the Galactic bulge, a dedicated near-IR spectroscopic follow-up is still necessary. \textit{Gaia}NIR is a proposed space mission that will map the entire sky in the NIR, extending the capabilities and deepness of \textit{Gaia}; if accepted, it will be launched in the 2040s, extending the baseline by 20 years, which translates into a $20\times$ more accurate proper motions.
The future ground and space facilities might change that scenario. The ELT will reach $\mu$as yr$^{-1}$ precision in proper motions \citep{2010MNRAS.402.1126T, 2021JATIS...7c5005R}, and there are JWST observations focusing on the Galactic bulge. The Wide-Field Instrument on board the Roman telescope will be ideal for NIR surveys with precise photometry (PSF FWHM $<0\farcs1$ ) and low resolution ($R\sim600$) multi-object spectroscopy, sufficient for radial velocity measurements. The Rubin observatory will map the southern sky and will reach proper motion precision of $0.2(1)$ mas yr$^{-1}$ for sources with magnitude $r=21(24)$ mag\footnote{\href{https://www.lsst.org/sites/default/files/docs/sciencebook/SB_3.pdf}{https://www.lsst.org/sites/default/files/docs/sciencebook/SB\_3.pdf}}.

\section*{Acknowledgements}

A.\,L. acknowledges support from the ANID Doctorado Nacional 2021 scholarship 21211520, and the ESO studentship. 
N.\,W.\,C.\,L. gratefully acknowledges the generous support of a Fondecyt Regular grant 1230082, as well as support from Millenium Nucleus NCN19\_058 (TITANs) and funding via the BASAL Centro de Excelencia en Astrofisica y Tecnologias Afines (CATA) grant PFB-06/2007. N.\,W.\,C.\,L. also thanks for support from ANID BASAL project ACE210002 and ANID BASAL projects ACE210002 and FB210003.
J.\,A.-G. acknowledges support from Fondecyt Regular 1201490 and ANID – Millennium Science Initiative Program – ICN12\_009 awarded to the Millennium Institute of Astrophysics MAS.
A.\,V.\ N. acknowledges support from the National Agency for Research and Development (ANID), Scholarship Program Doctorado Nacional 2020 – 21201226, ANID, Millennium Science Initiative, ICN12\_009 and ANID BASAL FB210003.
D.\,M. gratefully acknowledges support by the ANID BASAL projects ACE210002 and FB210003, by Fondecyt Project No. 1220724, and by CNPq/Brazil through project 350104/2022-0.
We gratefully acknowledge the use of data from the ESO Public Survey program IDs 179.B-2002 and 198.B-2004 taken with the VISTA telescope and data products from the Cambridge Astronomical Survey Unit.
This research made use of Astropy, a community-developed core Python package for Astronomy \citep{2018AJ....156..123A, 2013A&A...558A..33A}.
This research made use of NumPy \citep{harris2020array}, SciPy \citep{Virtanen_2020} and Scikit-learn \citep{scikit-learn}.

\section*{Data Availability}

We use data from the ESO Public Survey program IDs 179.B-2002 and 198.B-2004 taken with the VISTA telescope and data products from the Cambridge Astronomical Survey Unit. The VVV reduced images are available in the ESO Science Archive.
This work has made use of data from the European Space Agency (ESA) mission
{\it Gaia} (\url{https://www.cosmos.esa.int/gaia}), processed by the {\it Gaia}
Data Processing and Analysis Consortium (DPAC,
\url{https://www.cosmos.esa.int/web/gaia/dpac/consortium}). Funding for the DPAC
has been provided by national institutions, in particular, the institutions
participating in the {\it Gaia} Multilateral Agreement.
For the HVS candidates, we provide tables in the paper that will be available in CDS.




\bibliographystyle{mnras}
\bibliography{hvs} 




\appendix

\section{Possible \rev{M-dwarfs} in the \gaia DR3 -- VIRAC2 crossmatch}
\label{sec:appendixBDS_V2GDR3}


In this section, we present the list of the possible \rev{M-dwarfs} in our \gaia DR3 -- VIRAC2 sample (Table \ref{tab:bds_v2gdr3}) and in the VIRAC2 sample without \gaia DR3 counterparts (Table \ref{tab:bds_v2}) . These stars are selected with the colour cuts of Eq. \ref{eq:bds}, as described in Sec. \ref{sec:selection of stars}.

\begin{table*}
    \centering
    \begin{tabular}{lccccccc}
    \hline
    \gaia DR3 source ID & RA [hms] & Dec [dms] & $H$ [mag] & $J$ [mag] & $K_s$ [mag] & $\mu_{RA}$ [$mas\,yr^{-1}$]  & $\mu_{Dec}$ [$mas\,yr^{-1}$]  \\
        \hline
        \hline
4062689730907384576 & 17h59m46.35s & -28d17m16.22s & 13.77 & 14.41 & 13.58 & -13.8 $\pm$ 0.25 & -14.39 $\pm$ 0.25 \\
4056600017090916864 & 17h53m44.06s & -29d05m13.19s & 13.67 & 14.28 & 13.45 & 3.21 $\pm$ 1.45 & -36.8 $\pm$ 1.45 \\
4043668347447008512 & 17h54m29.02s & -31d27m13.07s & 13.69 & 14.37 & 13.47 & -23.66 $\pm$ 0.57 & -5.78 $\pm$ 0.57 \\
        \hline
    \end{tabular}
    \caption{Properties of the stars, in the \gaia DR3 -- VIRAC2 crossmatch, that could be \rev{M-dwarfs} according to the colour cuts of Eq. \ref{eq:bds}. The proper motions are from VIRAC2, without the reflex motion correction.}
    \label{tab:bds_v2gdr3}
\end{table*}

\begin{table*}
    \centering
    \begin{tabular}{ccccccc}
    \hline
    RA [hms] & Dec [dms] & $H$ [mag] & $J$ [mag] & $K_s$ [mag] & $\mu_{RA}$ [$mas\,yr^{-1}$]  & $\mu_{Dec}$ [$mas\,yr^{-1}$]  \\
        \hline
        \hline
17h54m40.43s & -29d36m24.30s & 12.98 & 13.63 & 12.76 & 11.97 $\pm$ 0.3 & 15.08 $\pm$ 0.3 \\
17h53m44.06s & -29d05m13.19s & 13.67 & 14.28 & 13.45 & 3.21 $\pm$ 1.45 & -36.8 $\pm$ 1.45 \\
17h58m39.42s & -28d45m03.90s & 13.6 & 14.21 & 13.41 & -1.19 $\pm$ 0.35 & -46.27 $\pm$ 0.35 \\
18h00m13.95s & -28d37m12.03s & 13.83 & 14.44 & 13.61 & -12.73 $\pm$ 0.49 & -15.64 $\pm$ 0.49 \\
17h54m10.59s & -28d51m53.11s & 13.7 & 14.28 & 13.47 & -4.12 $\pm$ 1.42 & -37.53 $\pm$ 1.42 \\
18h00m56.51s & -28d26m31.33s & 13.79 & 14.4 & 13.58 & 10.95 $\pm$ 0.38 & -12.65 $\pm$ 0.38 \\
18h03m07.61s & -27d19m08.33s & 13.4 & 14.05 & 13.25 & 16.96 $\pm$ 0.27 & -14.07 $\pm$ 0.27 \\
17h55m37.24s & -29d48m07.30s & 13.75 & 14.44 & 13.54 & 4.44 $\pm$ 0.31 & -23.15 $\pm$ 0.31 \\
17h54m29.02s & -31d27m13.07s & 13.69 & 14.37 & 13.47 & -23.66 $\pm$ 0.57 & -5.78 $\pm$ 0.57 \\
        \hline
    \end{tabular}
    \caption{Properties of the stars, in VIRAC2 without \gaia DR3 counterpart, that could be \rev{M-dwarfs} according to the colour cuts of Eq. \ref{eq:bds}. The proper motions are from VIRAC2, without the reflex motion correction.}
    \label{tab:bds_v2}
\end{table*}

\section{Stars in VIRAC2 pointing away from the GC}
\label{ap:table_virac2notGDR3_frmctr}

This section presents the photometric properties and derived parameters of the \rev{65} stars that have VIRAC2 data, but no \gaia DR3 counterpart, and whose VIRAC2 proper motions point away from $0.5\deg$ from the GC.

\LTcapwidth=\textwidth
\onecolumn
\begin{longtable}{rcccccc}

\hline
ID & RA [hms] & Dec [dms] & l [deg] & b [deg] & $K_s$ [mag] & $J-K_s$ [mag] ]\\
\hline
\hline
\endfirsthead

\hline
ID & RA [hms] & Dec [dms] & l [deg] & b [deg] & $K_s$ [mag] & $J-K_s$ [mag] ]\\
\hline
\hline
\endhead

\hline
\endfoot

\endlastfoot
1 & 17h51m54.16s & -28d01m38.73s & 1.4904 & -0.7146 & 14.28 & 2.26 \\
2 & 17h49m31.40s & -28d59m43.91s & 0.3911 & -0.7613 & 14.2 & 2.47 \\
3 & 17h48m40.41s & -29d08m22.46s & 0.172 & -0.6762 & 13.76 & 2.6 \\
4 & 17h51m11.57s & -28d04m06.82s & 1.375 & -0.6009 & 13.9 & 2.17 \\
5 & 17h51m33.79s & -27d03m47.14s & 2.2813 & -0.1578 & 14.83 & 3.73 \\
6 & 17h48m37.10s & -28d56m04.77s & 0.3413 & -0.5602 & 14.52 & 3.27 \\
7 & 17h46m56.33s & -26d52m27.36s & 1.913 & 0.8229 & 13.92 & 1.8 \\
8 & 17h46m20.34s & -28d08m31.05s & 0.7604 & 0.2784 & 14.45 & 2.85 \\
9 & 17h45m13.27s & -28d31m47.60s & 0.3011 & 0.2864 & 14.72 & 3.67 \\
10 & 17h42m02.43s & -27d46m11.88s & 0.5804 & 1.2819 & 14.0 & 1.73 \\
11 & 17h46m06.37s & -27d50m16.34s & 0.9934 & 0.4805 & 14.61 & 3.15 \\
12 & 17h46m40.54s & -27d40m40.52s & 1.1956 & 0.4559 & 14.86 & 3.96 \\
13 & 17h41m25.95s & -26d49m12.87s & 1.3162 & 1.8978 & 14.14 & 1.99 \\
14 & 17h42m14.52s & -27d48m39.42s & 0.5691 & 1.2225 & 14.02 & 1.92 \\
15 & 17h45m18.18s & -28d00m29.66s & 0.7556 & 0.5432 & 14.61 & 3.06 \\
16 & 17h43m10.80s & -29d03m55.52s & -0.3893 & 0.387 & 15.05 & 3.86 \\
17 & 17h41m30.29s & -29d16m18.41s & -0.7574 & 0.589 & 15.17 & 4.22 \\
18 & 17h40m51.43s & -29d53m26.62s & -1.3571 & 0.381 & 14.47 & 2.58 \\
19 & 17h42m56.86s & -28d47m26.55s & -0.1823 & 0.5747 & 14.41 & 3.63 \\
20 & 17h40m42.81s & -29d54m45.52s & -1.3922 & 0.3958 & 14.4 & 2.34 \\
21 & 17h42m12.50s & -28d57m57.80s & -0.4166 & 0.62 & 14.58 & 2.79 \\
22 & 17h28m51.23s & -32d26m46.33s & -4.8958 & 1.1563 & 14.39 & 2.3 \\
23 & 17h44m12.97s & -28d58m25.90s & -0.1925 & 0.2422 & 15.17 & 4.25 \\
24 & 17h41m47.24s & -29d51m26.34s & -1.2221 & 0.2276 & 14.79 & 3.58 \\
25 & 17h40m18.47s & -30d17m17.94s & -1.7572 & 0.2709 & 14.39 & 3.05 \\
26 & 17h37m08.53s & -30d54m12.78s & -2.6411 & 0.5185 & 14.7 & 3.06 \\
27 & 17h43m48.93s & -28d41m41.24s & -0.0008 & 0.4631 & 14.52 & 3.06 \\
28 & 17h34m24.30s & -30d25m27.57s & -2.5551 & 1.2727 & 13.89 & 1.36 \\
29 & 17h38m49.22s & -29d59m14.81s & -1.6737 & 0.7033 & 14.21 & 1.99 \\
30 & 17h35m55.31s & -28d38m22.47s & -0.8743 & 1.9593 & 13.83 & 1.35 \\
31 & 17h44m04.72s & -29d02m36.47s & -0.2676 & 0.2313 & 14.76 & 3.7 \\
32 & 17h44m04.69s & -29d09m25.08s & -0.3643 & 0.172 & 15.01 & 3.84 \\
33 & 17h40m15.79s & -29d57m06.96s & -1.4773 & 0.4577 & 14.38 & 2.61 \\
34 & 17h40m56.25s & -29d10m23.60s & -0.7392 & 0.7462 & 14.71 & 3.09 \\
35 & 17h43m29.07s & -29d13m15.07s & -0.4866 & 0.2488 & 15.01 & 3.97 \\
36 & 17h40m17.31s & -30d23m30.21s & -1.8471 & 0.2196 & 14.62 & 3.43 \\
37 & 17h32m26.70s & -29d35m51.44s & -2.0918 & 2.0782 & 13.84 & 1.55 \\
38 & 17h40m49.56s & -29d05m33.27s & -0.6837 & 0.8096 & 14.5 & 2.55 \\
39 & 17h43m04.90s & -29d48m44.04s & -1.0359 & 0.0127 & 15.2 & 4.31 \\
40 & 17h41m51.13s & -29d41m08.00s & -1.0688 & 0.3063 & 15.06 & 3.91 \\
41 & 17h45m47.85s & -28d57m43.36s & -0.0019 & -0.0466 & 14.58 & 4.74 \\
42 & 17h45m40.89s & -30d57m20.09s & -1.717 & -1.0633 & 14.51 & 2.73 \\
43 & 17h43m56.25s & -30d27m16.10s & -1.4854 & -0.4819 & 14.84 & 3.73 \\
44 & 17h44m41.96s & -30d50m34.58s & -1.7307 & -0.8248 & 15.08 & 4.06 \\
45 & 17h53m36.27s & -30d32m25.42s & -0.4863 & -2.311 & 13.4 & 1.24 \\
46 & 17h43m22.40s & -30d52m13.52s & -1.9032 & -0.5967 & 14.97 & 3.91 \\
47 & 17h44m57.45s & -30d50m45.38s & -1.7044 & -0.8736 & 14.48 & 3.17 \\
48 & 17h47m09.07s & -32d46m43.90s & -3.1142 & -2.2762 & 13.89 & 1.46 \\
49 & 17h45m27.22s & -30d08m00.62s & -1.0408 & -0.5931 & 14.99 & 3.89 \\
50 & 17h48m34.48s & -29d33m48.56s & -0.2024 & -0.8763 & 14.21 & 2.09 \\
51 & 17h43m16.85s & -32d48m15.56s & -3.5599 & -1.5952 & 13.98 & 1.52 \\
52 & 17h44m50.13s & -30d55m19.86s & -1.7831 & -0.8911 & 14.64 & 3.34 \\
53 & 17h46m07.04s & -30d51m52.92s & -1.5907 & -1.0958 & 14.29 & 2.23 \\
54 & 17h45m27.17s & -29d12m17.06s & -0.2482 & -0.1088 & 14.44 & 4.14 \\
55 & 17h46m33.01s & -29d25m09.19s & -0.3069 & -0.4247 & 14.96 & 3.7 \\
56 & 17h46m02.56s & -30d00m09.59s & -0.8627 & -0.6338 & 14.97 & 3.82 \\
57 & 17h44m45.93s & -30d20m20.38s & -1.2936 & -0.5736 & 15.02 & 3.83 \\
58 & 17h46m17.57s & -30d41m58.40s & -1.43 & -1.0422 & 14.4 & 3.23 \\
59 & 17h47m08.69s & -29d22m38.78s & -0.204 & -0.5137 & 14.81 & 3.47 \\
60 & 17h38m42.34s & -33d15m48.34s & -4.4571 & -1.0246 & 14.14 & 1.77 \\
61 & 17h47m50.76s & -29d41m50.15s & -0.3987 & -0.8098 & 14.06 & 2.77 \\
62 & 17h47m18.44s & -30d47m53.38s & -1.4012 & -1.2797 & 14.26 & 2.13 \\
63 & 17h46m57.49s & -29d41m40.51s & -0.4963 & -0.6434 & 14.62 & 3.36 \\
64 & 17h46m36.58s & -30d34m58.63s & -1.2949 & -1.0399 & 14.39 & 2.66 \\
65 & 17h42m19.02s & -31d10m41.50s & -2.2841 & -0.5662 & 14.7 & 3.23 \\
\hline
\caption{Location and photometric properties of the stars with no \gaia DR3 counterpart and VIRAC2 proper motions pointing away from $0.5\deg$ around the GC. The Galactic longitude and latitude have a precision of $0\farcs5$.\label{tab:frmctr_v}} \\
\end{longtable}

\begin{longtable}{rccccccc}

\hline
ID &  $\mu$ [$mas\,yr^{-1}$] & d [kpc] & $r_{GC}$ [kpc] & $V_{tan}$ [\kms] & $t_f$ [yr] & Approach [deg]\\
\hline
\hline
\endfirsthead

\hline
ID &  $\mu$ [$mas\,yr^{-1}$] & d [kpc] & $r_{GC}$ [kpc] & $V_{tan}$ [\kms] & $t_f$ [yr] & Approach [deg] & P\\
\hline
\hline
\endhead

\hline
\endfoot

\endlastfoot
1 & 19.08 & 11.64 & 3.53 & 1072.9 & 1362189 & 0.344 & 1.0 \\
2 & 19.99 & 10.07 & 1.96 & 982.98 & 68908 & 0.001 & 0.97 \\
3 & 26.28 & 7.98 & 0.17 & 1011.28 & 489957 & 0.234 & 0.86 \\
4 & 19.82 & 8.9 & 0.81 & 839.35 & 1213281 & 0.383 & 0.89 \\
5 & 16.6 & 10.83 & 2.73 & 856.75 & 2020771 & 0.496 & 1.0 \\
6 & 17.33 & 10.1 & 1.98 & 830.76 & 567325 & 0.293 & 0.98 \\
7 & 20.07 & 10.06 & 1.96 & 978.06 & 1673409 & 0.32 & 0.97 \\
8 & 22.83 & 10.09 & 1.97 & 1128.96 & 610088 & 0.069 & 1.0 \\
9 & 21.35 & 10.17 & 2.05 & 1044.98 & 32379 & 0.279 & 1.0 \\
10 & 16.58 & 11.0 & 2.89 & 871.98 & 124411 & 0.041 & 1.0 \\
11 & 19.51 & 10.6 & 2.48 & 1021.4 & 899316 & 0.06 & 1.0 \\
12 & 16.93 & 10.5 & 2.38 & 863.32 & 1119425 & 0.26 & 0.92 \\
13 & 31.41 & 11.08 & 2.99 & 1683.19 & 1483537 & 0.055 & 1.0 \\
14 & 18.24 & 10.77 & 2.65 & 947.92 & 1136668 & 0.31 & 1.0 \\
15 & 28.29 & 10.56 & 2.44 & 1418.33 & 62984 & 0.228 & 1.0 \\
16 & 14.64 & 11.7 & 3.58 & 823.63 & 516474 & 0.225 & 0.93 \\
17 & 16.33 & 11.5 & 3.39 & 899.95 & 854837 & 0.122 & 1.0 \\
18 & 17.4 & 11.61 & 3.5 & 979.57 & 1216416 & 0.108 & 1.0 \\
19 & 21.66 & 9.27 & 1.15 & 960.71 & 466379 & 0.018 & 0.98 \\
20 & 17.45 & 10.77 & 2.65 & 906.58 & 1247434 & 0.358 & 0.98 \\
21 & 14.3 & 11.39 & 3.27 & 774.61 & 711158 & 0.209 & 0.95 \\
22 & 14.27 & 10.75 & 2.76 & 727.1 & 4794263 & 0.034 & 0.93 \\
23 & 15.41 & 10.82 & 2.7 & 794.32 & 283765 & 0.033 & 0.93 \\
24 & 14.52 & 11.37 & 3.26 & 788.06 & 1174594 & 0.385 & 0.92 \\
25 & 19.44 & 9.72 & 1.62 & 898.35 & 1451589 & 0.034 & 1.0 \\
26 & 13.51 & 11.76 & 3.67 & 752.89 & 2636652 & 0.126 & 0.98 \\
27 & 22.91 & 10.14 & 2.01 & 1112.38 & 348364 & 0.206 & 1.0 \\
28 & 16.36 & 10.86 & 2.78 & 853.39 & 2540906 & 0.297 & 0.94 \\
29 & 15.64 & 10.76 & 2.66 & 800.31 & 1652376 & 0.267 & 0.95 \\
30 & 18.3 & 10.53 & 2.44 & 917.74 & 1805352 & 0.056 & 1.0 \\
31 & 25.02 & 10.35 & 2.23 & 1240.06 & 254584 & 0.068 & 1.0 \\
32 & 28.88 & 11.31 & 3.19 & 1576.01 & 269883 & 0.251 & 1.0 \\
33 & 16.68 & 11.32 & 3.21 & 900.49 & 1363378 & 0.089 & 1.0 \\
34 & 15.38 & 11.51 & 3.39 & 850.99 & 964212 & 0.252 & 0.96 \\
35 & 15.42 & 10.93 & 2.81 & 800.31 & 500993 & 0.311 & 0.97 \\
36 & 15.99 & 10.6 & 2.5 & 807.72 & 1674628 & 0.313 & 0.94 \\
37 & 21.79 & 10.57 & 2.49 & 1127.72 & 2274099 & 0.143 & 1.0 \\
38 & 14.32 & 11.3 & 3.19 & 770.23 & 1008236 & 0.007 & 0.93 \\
39 & 13.14 & 11.76 & 3.64 & 733.39 & 1028925 & 0.407 & 0.92 \\
40 & 15.15 & 10.98 & 2.87 & 791.94 & 1028237 & 0.19 & 0.96 \\
41 & 47.89 & 8.24 & 0.12 & 1875.12 & 24273 & 0.027 & 1.0 \\
42 & 17.44 & 11.84 & 3.73 & 983.76 & 1741107 & 0.137 & 1.0 \\
43 & 18.43 & 10.56 & 2.45 & 925.84 & 1309472 & 0.316 & 1.0 \\
44 & 17.11 & 11.39 & 3.29 & 929.66 & 1668731 & 0.028 & 1.0 \\
45 & 30.39 & 9.15 & 1.09 & 1325.27 & 1542188 & 0.176 & 1.0 \\
46 & 20.7 & 10.38 & 2.28 & 1022.51 & 1578184 & 0.058 & 1.0 \\
47 & 21.18 & 9.97 & 1.87 & 1024.82 & 1498155 & 0.03 & 1.0 \\
48 & 15.51 & 10.27 & 2.23 & 754.23 & 3526367 & 0.411 & 0.94 \\
49 & 25.96 & 11.41 & 3.3 & 1413.54 & 846379 & 0.183 & 1.0 \\
50 & 17.76 & 10.54 & 2.42 & 893.5 & 768377 & 0.245 & 1.0 \\
51 & 14.75 & 10.78 & 2.73 & 753.78 & 3656984 & 0.128 & 0.95 \\
52 & 20.11 & 10.97 & 2.87 & 1079.99 & 1600298 & 0.496 & 1.0 \\
53 & 20.14 & 10.84 & 2.74 & 1041.09 & 1549381 & 0.231 & 1.0 \\
54 & 39.51 & 8.12 & 0.04 & 1520.99 & 155229 & 0.226 & 1.0 \\
55 & 23.87 & 10.71 & 2.59 & 1234.92 & 386132 & 0.264 & 1.0 \\
56 & 14.22 & 11.37 & 3.25 & 766.62 & 1021822 & 0.0 & 0.99 \\
57 & 16.58 & 10.62 & 2.51 & 836.06 & 1250916 & 0.396 & 1.0 \\
58 & 17.12 & 9.89 & 1.79 & 803.86 & 1539368 & 0.265 & 0.95 \\
59 & 14.31 & 11.28 & 3.16 & 769.73 & 526093 & 0.105 & 0.93 \\
60 & 15.52 & 11.42 & 3.39 & 848.9 & 4179327 & 0.357 & 0.99 \\
61 & 23.63 & 9.27 & 1.16 & 1073.24 & 668479 & 0.096 & 1.0 \\
62 & 19.26 & 11.06 & 2.96 & 1036.36 & 155673 & 0.481 & 1.0 \\
63 & 18.39 & 10.47 & 2.35 & 938.49 & 682123 & 0.466 & 0.97 \\
64 & 15.11 & 10.79 & 2.68 & 774.97 & 1538132 & 0.237 & 0.93 \\
65 & 23.2 & 11.24 & 3.15 & 1274.34 & 1758733 & 0.06 & 1.0 \\
\hline
\caption{Derived parameters of the stars with no \gaia DR3 counterpart and VIRAC2 proper motions pointing out from $0.5\deg$ around the GC. The second to last column indicates how close the projected 2D velocity vector approaches the GC. \rev{The last column is the probability of exceeding the escape velocity} \label{tab:frmctr_v2}} \\
\end{longtable}

\twocolumn


\bsp	
\label{lastpage}
\end{document}